\documentclass{article} %
\usepackage[T1]{fontenc}
\usepackage{iclr2027_conference,times}

\usepackage{amsmath,amsfonts,bm}

\def\eqref#1{equation~\ref{#1}}

\def\1{\bm{1}}

\DeclareMathAlphabet{\mathsfit}{\encodingdefault}{\sfdefault}{m}{sl}
\SetMathAlphabet{\mathsfit}{bold}{\encodingdefault}{\sfdefault}{bx}{n}

\usepackage{hyperref}
\hypersetup{hidelinks,
pdftitle={ACARe: Agent-Context-Aware Reranker for Iterative Search},
pdfauthor={Sahel Sharifymoghaddam, Yauhen Babakhin, Reza Esfandiarpoor, Oliver Holworthy, Jimmy Lin, Bo Liu},
pdfsubject={Preprint. Work in progress. Model, training data, and code will be released soon.}}
\usepackage{url}

\usepackage{booktabs}
\usepackage{xcolor}
\usepackage{wrapfig}
\usepackage{graphicx}   %
\usepackage[section]{placeins}
\usepackage{rotating}   %
\usepackage{cleveref}
\newsavebox{\appendixtablebox}
\newcommand{\fitappendixtable}[1]{%
  \sbox{\appendixtablebox}{\fontsize{8}{9.5}\selectfont
    \renewcommand{\scriptsize}{\fontsize{6}{7}\selectfont}%
    \renewcommand{\std}[1]{{\fontsize{6}{7}\selectfont($\pm$##1)}}%
    \renewcommand{\gain}[1]{{\fontsize{6}{7}\selectfont\color{gaingreen}(+##1)}}%
    \renewcommand{\loss}[1]{{\fontsize{6}{7}\selectfont\color{lossred}(--##1)}}%
    \renewcommand{\nodelta}{{\fontsize{6}{7}\selectfont\color{nochange}(--)}}%
    #1}%
  \ifdim\wd\appendixtablebox>\linewidth
    \PackageWarning{paper-tables}{Table body exceeds available width; font scaling applied}%
    \resizebox{\linewidth}{!}{\usebox{\appendixtablebox}}%
  \else\usebox{\appendixtablebox}\fi}
\newcommand{\std}[1]{{\scriptsize($\pm$#1)}}
\definecolor{gaingreen}{HTML}{157F3B}
\definecolor{lossred}{HTML}{C0392B}
\definecolor{nochange}{HTML}{8A8A8A}
\newcommand{\gain}[1]{{\scriptsize\color{gaingreen}(+#1)}}
\newcommand{\loss}[1]{{\scriptsize\color{lossred}(--#1)}}
\newcommand{\nodelta}{{\scriptsize\color{nochange}(--)}}
\newcommand{\method}{ACARe}
\makeatletter
\newenvironment{inplacetable}{\par\addvspace{12pt}\noindent\begin{minipage}{\linewidth}\def\@captype{table}}{\end{minipage}\par\addvspace{12pt}}
\makeatother
\usepackage{mdframed}

\title{ACARe: Agent-Context-Aware Reranker for Iterative Search}

\renewcommand{\thefootnote}{\fnsymbol{footnote}}

\author{
\textbf{Sahel Sharifymoghaddam}\textsuperscript{1}\thanks{Work done during an internship at NVIDIA.} \quad
\textbf{Yauhen Babakhin}\textsuperscript{2} \quad
\textbf{Reza Esfandiarpoor}\textsuperscript{2} \\
\textbf{Oliver Holworthy}\textsuperscript{2} \quad
\textbf{Jimmy Lin}\textsuperscript{1} \quad
\textbf{Bo Liu}\textsuperscript{2} \\[0.6em]
\textsuperscript{1}University of Waterloo, Canada \\
\textsuperscript{2}NVIDIA \\[0.3em]
\texttt{\{sahel.sharifymoghaddam,jimmylin\}@uwaterloo.ca} \\
\texttt{\{ybabakhin,
resfandiarpo,oholworthy,boli\}@nvidia.com}
}


\iclrfinalcopy

\begin{document}
\raggedbottom

\maketitle
\lhead{Preprint. Work in progress.}
\renewcommand{\thefootnote}{\arabic{footnote}}
\begin{abstract}
\looseness=-1
In agentic search, a document can match an agent-generated sub-query without helping answer the user's original question.
Conventional rerankers score documents against the sub-query alone, overlooking the broader task and the agent's reasoning.
We introduce the Agent-Context-Aware Reranker (\method{}), a compact 4B LLM trained to rerank documents using this context when available.
Across multiple benchmarks and agent families, \method{} improves retrieval quality over its base reranker while reducing search effort.
These benefits extend to answer generation: \method{} improves accuracy and reduces searches over no reranking, while matching or exceeding the base reranker's mean accuracy with more precise evidence.
It also helps smaller agents narrow the retrieval-quality gap to larger agents, showing how better retrieval tools can make smaller agents more effective.
The model, training data, and code will be released soon.
\end{abstract}

\section{Introduction}\label{sec:intro}

Search agents answer complex questions by searching for evidence over multiple steps~\citep{yao2023react,jin2025search,tongyidr2025,li2026openresearcher}.
Given a user's question, an agent issues a search query, reads the returned documents, and decides what to search for next.
We call each agent-generated search query a \emph{sub-query}.
A reranker scores the retrieved documents to prioritize the evidence the agent sees.
Conventional rerankers score each document against only the sub-query, which is often keyword-like~\citep{meng2026revisiting} and omits the broader task.
A document can therefore match a sub-query without helping answer the user's original question.

The original question states the agent's goal, while its available reasoning trace explains what evidence is needed next.
Neither requires additional generation.
Some first-stage retrievers~\citep{chen2026agentir,chen2026iter} use this context, but deploying them requires local access to the document corpus.
A second-stage reranker, however, refines candidates from any source, supporting lexical and dense retrieval as well as search APIs exposing document text without corpus access, where context-aware first-stage retrieval is infeasible.

We introduce the Agent-Context-Aware Reranker (\method{}), a 4B decoder-only LLM fine-tuned from \href{https://huggingface.co/Qwen/Qwen3-Reranker-4B}{Qwen3-Reranker-4B}~\citep{zhang2025qwen3embed}, a compact, strong open-weight reranker.
We use the sub-query and agent context to score candidates independently, which avoids the sequential sliding-window passes that listwise rerankers require for long candidate lists~\citep{sun2023rankgpt,pradeep2023rankvicuna}.
To train \method{}, we combine new agent trajectories with original and re-mined DR-Synth~\citep{chen2026agentir} labels.
Re-mining changes candidates but preserves sub-queries and traces.
We incorporate independent context dropout to support missing fields and proprietary agents that do not expose reasoning traces.
We also show that \method{}'s benefits combine with context-aware first-stage retrieval (\Cref{sec:agentic-results}), complementing existing retrievers.
In summary, our contributions are:
\begin{itemize}
\item \textbf{A context-aware LLM reranker for agentic document search.}
To our knowledge, we introduce the first pointwise LLM reranker trained for search-agent sub-queries using agent context.
Our context formulation and context dropout let one checkpoint rerank with the original question, reasoning trace, both, or neither, including for proprietary agents that hide their reasoning.
\item \textbf{Expanded training data.}
We extend DR-Synth with original questions, re-mined candidates, and a new rollout, creating 16{,}601 labeled instances for context-aware reranking.
\item \textbf{Retrieval and answer-generation gains.}
Across three benchmarks, four first-stage retrievers plus API-based search, and two harnesses, \method{} improves dataset-averaged nDCG@10 over the base reranker by 0.9--2.0 percentage points (pp), with fewer searches for all five main agents.
In end-to-end answer generation on ClimbMix~\citep{diao2025nemotron}, it raises accuracy by 8.2--9.1 pp over no reranking and matches or exceeds the base reranker's mean accuracy.
In the BCP context control, \method{}'s context formulation drives most of its gain over the base reranker even with the base model unchanged, while training yields the best mean on every metric.
\end{itemize}

\section{Background and related work}\label{sec:related}

\paragraph{Retrieval for search agents.}
Search agents interleave reasoning and external actions; deep-research systems extend this pattern to sustained search and evidence synthesis~\citep{yao2023react,jin2025search,tongyidr2025,li2026openresearcher}.
Fixed-corpus benchmarks such as BrowseComp-Plus enable reproducible evaluation of these interactions~\citep{chen-etal-2026-browsecomp}.
Unlike a user question, an agent's sub-query is generated within a trajectory and may omit information already present in the conversation.
\citet{meng2026revisiting} find that these queries often consist of keywords, phrases, and exact-match expressions, differing from the natural-language questions commonly used to train neural rankers.
This motivates retrieval methods that use the agent's context.

\paragraph{Context-aware first-stage retrieval.}
AgentIR embeds the reasoning trace with the sub-query~\citep{chen2026agentir}, while ITER also incorporates the original question and preceding sub-queries~\citep{chen2026iter}.
Both require local access to the document corpus, whereas \method{} reranks returned documents.

\paragraph{Session and conversational search.}
Context-aware ranking predates search agents: session-search models rank documents using a user's earlier in-session search activity~\citep{ahmad2019context}, and conversational search resolves context-dependent questions by rewriting them or by conditioning retrieval and reranking on dialogue history~\citep{dalton2020cast19,anantha2021qrecc,qu2020orconvqa}.
\method{} instead conditions a model-generated, often keyword-like sub-query on the agent's original question and its own reasoning trace, with one checkpoint handling either field's absence.

\paragraph{Reasoning-intensive retrieval.}
ReasonIR, RaDeR, and ReasonEmbed train dense retrievers~\citep{shao2025reasonir,das-etal-2025-rader,chen-etal-2026-reasonembed}.
Their focus is improving one-shot neural retrieval rather than interpreting sub-queries within an evolving agent trajectory.

\paragraph{Retrieval learning from agent feedback.}
LRAT learns from the agent's browsing actions, unbrowsed candidates, and post-browse reasoning~\citep{zhou2026lrat}.
Agentic-R combines local query--passage relevance with global answer correctness to measure passage utility and alternates agent and retriever optimization~\citep{liu2026agenticr}.
Critic-R uses post-retrieval reasoning to refine queries and converts feedback from successful and failed refinements into retriever supervision~\citep{alam2026criticr}.
CoSearch jointly trains an agent and a generative ranker using rewards that combine ranking quality with trajectory outcomes~\citep{zeng2026cosearch}.
These systems use downstream agent behavior to train the retrieval tool, sometimes jointly with the agent.
\method{} focuses on a different requirement: using the original question and available reasoning trace to interpret each sub-query at ranking time, with one checkpoint supporting all four context modes.

\paragraph{Reranking in agentic search.}
Pointwise and generative listwise rerankers refine first-stage candidates~\citep{zhang2025qwen3embed,sun2023rankgpt,pradeep2023rankvicuna}.
Instruction-aware retrieval includes a reranker conditioned on human-written task descriptions~\citep{asai-etal-2023-task}.
\citet{sharifymoghaddam-lin-2026-rerank} use the search agent's own LLM at low reasoning effort as a listwise reranker; medium-effort search with reranking can outperform high-effort search without reranking in retrieval quality and answer accuracy at lower effective token cost.
\citet{meng2026revisiting} report retrieval and answer-quality gains from reranking, generally with fewer searches, and test query-only and reasoning-conditioned sub-query rewriting.
SetR selects passage sets that satisfy a query's information requirements~\citep{lee2025setr}, while Rank4Gen learns selection and ordering from downstream generation preferences~\citep{fan2026rank4gen}.
Generative ranking and selection add autoregressive decoding overhead; long candidate lists can also require sequential sliding-window passes~\citep{sun2023rankgpt,pradeep2023rankvicuna}.
Concurrent to our work, RubricRanker~\citep{liu2026rubricranker} trains a generative listwise selector with search rubrics, supporting reasoning-conditioned search and question-only RAG.
\method{} instead scores documents independently and trains one checkpoint for four context modes, including question-only context when traces are hidden.

\section{\method{} Methodology}\label{sec:method}

\subsection{Problem formulation}\label{sec:prelim}

We consider an agent that gathers evidence for an original question $Q$ through a ReAct-style loop~\citep{yao2023react}.
Its output is either a ranked document list for an upstream orchestrator or an evidence-grounded answer for a deep-search task.
Both settings share the trajectory
\begin{equation}
    \mathcal{S} = (Q,\tau_1,a_1,o_1,\ldots,\tau_N,a_N,o_N,y),
    \label{eq:agent-trajectory}
\end{equation}
where $N$ is the number of tool-interaction turns, $a_t$ is an action, $o_t$ its observation, and $\tau_t$ the preceding reasoning text, when available.
At turn $t$, the agent conditions its generation on the history $h_t=(Q,\tau_1,a_1,o_1,\ldots,\tau_{t-1},a_{t-1},o_{t-1})$.
A search action contains sub-queries $q_{t,1},\ldots,q_{t,n_t}$; retrieval and reranking supply the documents in $o_t$.
Sub-queries in the same turn share $\tau_t$, and we write $q_t$ for an individual sub-query when its within-turn index is unnecessary.

Let $\mathcal{D}_{\mathrm{seen}}$ denote the documents returned during the trajectory.
The terminal generation conditioned on $h_{N+1}$ has one of two forms:
\begin{equation}
    y =
    \begin{cases}
        (\operatorname{id}(d_{\pi(1)}),\ldots,\operatorname{id}(d_{\pi(K)})), & \text{retrieval-only},\\
        (w_1,\ldots,w_L), & \text{answer generation},
    \end{cases}
    \label{eq:agent-output}
\end{equation}
where $\operatorname{id}(d)$ is a document ID, $\pi$ orders $K$ distinct documents from $\mathcal{D}_{\mathrm{seen}}$, and $(w_1,\ldots,w_L)$ is an evidence-grounded answer.
Our reranker supports both outputs; the main experiments evaluate retrieval-only output, and a transfer experiment evaluates answer generation (\Cref{sec:piika}).

Conventional rerankers score candidates against $q_t$ alone, although sub-queries may omit the search intent or constraints in $Q$.
We include available agent context to recover this information.

\subsection{Agent context}\label{sec:context}

We use two context fields: the original question $Q$ and the reasoning trace $\tau_t$ preceding the current search.
The former is fixed throughout the session; the latter is the text the agent has already produced, when exposed by its interface.
We do not make an additional model call to elicit a trace.
Context therefore requires no additional agent generation, although including it increases the reranker's input length.
We define four input modes: $\mathrm{M0}$ contains $q_t$ alone; $\mathrm{M1}$ contains $(q_t,\tau_t)$; $\mathrm{M2}$ contains $(q_t,Q)$; and $\mathrm{M3}$ contains $(q_t,\tau_t,Q)$.
The instruction and query block depend on the mode, while the system message and document field are shared (\Cref{fig:prompt-modes}).
In $\mathrm{M0}$, the prompt is identical to the base Qwen3-Reranker prompt.
Following \citet{chen2026agentir}, we use the current trace rather than concatenating traces from earlier turns.

The decoder-only LLM jointly processes the sub-query, retained context, and one candidate document.
Relevance can therefore depend on the available agent context, without requiring each document to answer the whole question.
Context availability determines the inference mode.
Agents that expose their reasoning can use $\mathrm{M3}$ when $Q$ is also available.
Agents that do not expose reasoning can use $\mathrm{M2}$; deployments without either field use $\mathrm{M0}$.
The inference-context ablation in \Cref{sec:context-ablation} compares these modes for one agent and benchmark.

\subsection{Multi-task training over context modes}\label{sec:multitask}

\method{} is initialized from Qwen3-Reranker-4B and fine-tuned as a single checkpoint supporting all four modes.
During training, the reasoning trace and original question are independently dropped with probability $0.5$.
A dropped or unavailable field is omitted, and the prompt uses the instruction corresponding to the remaining fields.
When both fields are available, each mode has probability $0.25$.
Missing source fields can change the effective mode distribution.
Dropout decisions are deterministic functions of the seed, epoch, field, and query, ensuring consistent masks across workers.

Each training instance is a sub-query with its available context, one positive document, and retrieved negatives.
The same sampled context is used for all documents in its group.
The model scores each query--document pair using the difference between the next-token logits for \texttt{yes} and \texttt{no}.
Following \citet{10.1007/978-3-030-72240-1_26}, we use localized contrastive estimation (LCE) rather than pointwise binary cross-entropy to train the reranker to distinguish a positive document from retrieved negatives.
LCE applies cross-entropy over the group's scores with temperature $0.5$.
At inference, a sigmoid converts the logit difference to a relevance score without changing document order.
Training inputs are capped at 5{,}120 tokens so that eight query--document pairs (one training group), model state, and activations fit within each GPU's 80\,GB memory; inference uses an 8{,}192-token cap.
\Cref{app:training} gives the exact objective and optimization settings.

A single checkpoint thus supports both agentic and one-shot retrieval.
In the one-shot evaluation (\Cref{tab:oneshot-main}), it uses $\mathrm{M0}$, so comparisons with the base reranker measure the effect of fine-tuning without additional inference context.

\subsection{Training data}\label{sec:data}

Training requires relevance labels for an agent's sub-queries rather than only for the original question.
We extend the DR-Synth construction of \citet{chen2026agentir} to retain $Q$ alongside each sub-query and its reasoning trace.
Starting from questions with gold answers and supporting documents, an agent generates search trajectories.
For each sub-query, an oracle reranker receives the retrieved candidates, gold supporting documents, original question, and gold answer.
Its ranking supplies positive and negative documents for that sub-query.
A separate judge filters trajectories by comparing the agent's final answer with the gold answer.

We combine three label sets from two rollouts.
We retain the original DR-Synth labels and retrieve and label new candidates for its recorded sub-queries using \mbox{Nemotron-3-Embed-8B}\footnote{\url{https://huggingface.co/nvidia/Nemotron-3-Embed-8B-BF16}} as a second retriever.
Re-mining changes the supervision documents but preserves the sub-queries and reasoning traces.
We also generate a new rollout using that second retriever, with the oracle's top-ranked documents returned to the agent during search.
The pool contains 16{,}601 instances, versus 5{,}238 in DR-Synth.
We select the data mixture and hyperparameters on a validation split grouped by original question, then train the final checkpoint on the full pool.
\Cref{app:training} specifies the models, candidate construction, filtering, and size of each label set.

\section{Experimental setup}\label{sec:setup}

We use NVIDIA's Retrieval Bench\footnote{\url{https://github.com/NVIDIA/NeMo-Retriever/tree/main/retrieval-bench}} for iterative document retrieval and Piika\footnote{\url{https://github.com/castorini/piika}} for end-to-end answer generation.
Piika is a Pi-based deep-search harness that integrates RankLLM~\citep{sharifymoghaddam2025rankllm}\footnote{\url{https://github.com/castorini/rank_llm}} for reranking.
Both evaluation harnesses differ from the AgentIR harness used to collect the training trajectories (\Cref{app:training}).
Our retrieval evaluations cover BrowseComp-Plus (BCP; \citealp{chen-etal-2026-browsecomp}), BRIGHT~\citep{su2024bright}, and Total Recall QA (TRQA; \citealp{rafiee2026trqa}), using 200, 1{,}384, and 569 questions, respectively.
One-shot retrieval and Retrieval Bench report nDCG@10 and Recall@10 on the final document ranking; Retrieval Bench additionally measures search calls, LLM turns, and agent API cost per query.
Piika reports end-to-end answer accuracy, Recall@all, Precision@all, and searches per query on BrowseComp-Plus$_{\mathrm{CM}}$ (\citealp{sharifymoghaddam2026cmass}) and BCP (\Cref{sec:piika}).
In all our cost calculations, we exclude retrieval and reranker serving costs: the neural retrievers and rerankers used in the main agentic experiments are relatively small, locally hosted 4B--8B models, whose serving costs we assume are negligible compared with agent LLM API costs (\Cref{app:cost}).

BM25 provides the lexical baseline.
\mbox{AgentIR-4B} is designed for agentic retrieval and achieved the highest downstream accuracy among the first-stage retrievers compared in its BrowseComp-Plus deep-search evaluation~\citep{chen2026agentir}.
\mbox{Qwen3-Embedding-4B}~\citep{zhang2025qwen3embed} is its base model, providing a general-purpose dense baseline.
\mbox{Nemotron-3-Embed-8B} represents a leading retrieval model: it ranked first on the multilingual Retrieval Embedding Benchmark (RTEB)\footnote{\url{https://huggingface.co/blog/rteb}} in the July 2026 release report.\footnote{\url{https://huggingface.co/blog/nvidia/nemotron-3-embed-wins-rteb}}
Strong first-stage retrievers leave less headroom for reranking.
One-shot tests compare no reranking, Jina-Reranker-v3 (0.6B)~\citep{wang2025jinareranker3}, Qwen3-Reranker-4B, and \method{}-4B on the same candidates.
In this setting, \method{} uses the context-free mode $\mathrm{M0}$.
Unless noted, Qwen3-Reranker-4B receives only the sub-query in its native prompt.

Retrieval Bench connects an LLM agent to a search tool, manages iterative search, and records the ranked document IDs submitted through \texttt{final\_results}; \Cref{app:repro} explains the harness and serving details.
Each search normally retrieves 100 candidates, reranks them when enabled, and returns ten unseen documents.

The main Retrieval Bench experiments use five LLMs as agents: two open-weight models, Nemotron-3 Ultra and MiniMax M3, and three proprietary models, Gemini 3.6 Flash, GPT-5.6 Terra, and Claude Sonnet 5.
Throughout this paper, Nemotron-3 Ultra denotes the NVFP4 version.
The open-weight agents expose reasoning traces and use $\mathrm{M3}$; the proprietary agents use $\mathrm{M2}$, which requires only the sub-query and original question.
The three smaller proprietary models in the size--cost experiment (\Cref{sec:size-cost}) also use $\mathrm{M2}$.

Each configuration is evaluated five times on a fixed question sample, except the Piika studies, which use six runs.
We average quality metrics over benchmark subsets, then average the three benchmarks equally.
Tables report means and sample standard deviations across runs; quality metrics use a 0--100 scale.
Differences and percentage changes are computed before rounding, so they may differ from calculations using displayed values.
Bold identifies the best mean within each comparison group.
Symbols $\dagger$ and $*$ mark statistically significant differences from no reranking and Qwen3-Reranker-4B, respectively.
Unmarked results may be nonsignificant or untested; \Cref{app:stats} specifies the tested comparisons and correction families.

\section{Main results}

\begin{table}[!tp]
\centering
\setlength{\tabcolsep}{2.5pt}
\caption{\textbf{One-shot retrieval.}
$\Delta$: change from no reranking.
Group (e): BM25 on BCP, \mbox{Nemotron-3-Embed-8B} on BRIGHT, and \mbox{AgentIR-4B} on TRQA.}
\label{tab:oneshot-main}
\renewcommand{\arraystretch}{0.9}
\fitappendixtable{%
\begin{tabular}{@{}l llll llll@{}}
\toprule
\textbf{Retriever} & \multicolumn{4}{c}{\textbf{nDCG@10}\,$(\Delta)$} & \multicolumn{4}{c}{\textbf{Recall@10}\,$(\Delta)$} \\
\cmidrule(lr){2-5}\cmidrule(lr){6-9}
$+$\textbf{Reranker} & BCP & BRIGHT & TRQA & Avg. & BCP & BRIGHT & TRQA & Avg. \\
\midrule
(a1)~BM25 & 9.7~\nodelta & 14.1~\nodelta & 12.4~\nodelta & 12.0~\nodelta & 10.3~\nodelta & 16.4~\nodelta & 7.7~\nodelta & 11.5~\nodelta \\
(a2)~$+$~Jina-Reranker-v3 & 17.2~\gain{7.5} & 19.4~\gain{5.3} & 15.8~\gain{3.4} & 17.4~\gain{5.4} & 16.4~\gain{6.1} & 21.7~\gain{5.3} & 10.6~\gain{2.9} & 16.3~\gain{4.8} \\
(a3)~$+$~Qwen3-Reranker-4B & 21.3~\gain{11.6} & 21.6~\gain{7.5} & 17.3~\gain{4.9} & 20.1~\gain{8.1} & 19.1~\gain{8.8} & 24.4~\gain{8.0} & 11.7~\gain{4.0} & 18.4~\gain{6.9} \\
(a4)~$+$~\method{}-4B ($\mathrm{M0}$) & \textbf{23.3}~\gain{13.6} & \textbf{22.9}~\gain{8.8} & \textbf{19.1}~\gain{6.7} & \textbf{21.8}~\gain{9.8} & \textbf{20.6}~\gain{10.3} & \textbf{25.4}~\gain{9.0} & \textbf{12.2}~\gain{4.5} & \textbf{19.4}~\gain{7.9} \\
\midrule
(b1)~Qwen3-Embedding-4B & 12.9~\nodelta & 21.7~\nodelta & 9.8~\nodelta & 14.8~\nodelta & 14.7~\nodelta & 26.3~\nodelta & 7.1~\nodelta & 16.0~\nodelta \\
(b2)~$+$~Jina-Reranker-v3 & 23.6~\gain{10.7} & 21.8~\gain{0.1} & 14.3~\gain{4.5} & 19.9~\gain{5.1} & 26.1~\gain{11.4} & 25.5~\loss{0.8} & 10.6~\gain{3.5} & 20.7~\gain{4.7} \\
(b3)~$+$~Qwen3-Reranker-4B & 31.3~\gain{18.4} & 26.3~\gain{4.6} & 15.9~\gain{6.1} & 24.5~\gain{9.7} & 31.1~\gain{16.4} & 32.2~\gain{5.9} & 11.2~\gain{4.1} & 24.8~\gain{8.8} \\
(b4)~$+$~\method{}-4B ($\mathrm{M0}$) & \textbf{35.9}~\gain{23.0} & \textbf{28.6}~\gain{6.9} & \textbf{18.3}~\gain{8.5} & \textbf{27.6}~\gain{12.8} & \textbf{35.7}~\gain{21.0} & \textbf{34.4}~\gain{8.1} & \textbf{12.7}~\gain{5.6} & \textbf{27.6}~\gain{11.6} \\
\midrule
(c1)~AgentIR-4B & 26.7~\nodelta & 22.5~\nodelta & 17.6~\nodelta & 22.3~\nodelta & 30.2~\nodelta & 27.6~\nodelta & 11.8~\nodelta & 23.2~\nodelta \\
(c2)~$+$~Jina-Reranker-v3 & 31.7~\gain{5.0} & 22.1~\loss{0.4} & 17.4~\loss{0.2} & 23.7~\gain{1.4} & 35.7~\gain{5.5} & 26.4~\loss{1.2} & 12.5~\gain{0.7} & 24.9~\gain{1.7} \\
(c3)~$+$~Qwen3-Reranker-4B & 44.6~\gain{17.9} & 27.3~\gain{4.8} & 19.9~\gain{2.3} & 30.6~\gain{8.3} & 45.5~\gain{15.3} & 33.0~\gain{5.4} & 14.3~\gain{2.5} & 30.9~\gain{7.7} \\
(c4)~$+$~\method{}-4B ($\mathrm{M0}$) & \textbf{48.7}~\gain{22.0} & \textbf{29.4}~\gain{6.9} & \textbf{22.1}~\gain{4.5} & \textbf{33.4}~\gain{11.1} & \textbf{49.7}~\gain{19.5} & \textbf{34.9}~\gain{7.3} & \textbf{15.2}~\gain{3.4} & \textbf{33.3}~\gain{10.1} \\
\midrule
(d1)~Nemotron-3-Embed-8B & 28.0~\nodelta & 25.3~\nodelta & 14.4~\nodelta & 22.6~\nodelta & 31.3~\nodelta & 30.2~\nodelta & 10.2~\nodelta & 23.9~\nodelta \\
(d2)~$+$~Jina-Reranker-v3 & 30.8~\gain{2.8} & 22.1~\loss{3.2} & 16.0~\gain{1.6} & 22.9~\gain{0.3} & 33.2~\gain{1.9} & 26.2~\loss{4.0} & 11.9~\gain{1.7} & 23.7~\loss{0.2} \\
(d3)~$+$~Qwen3-Reranker-4B & 43.1~\gain{15.1} & 27.0~\gain{1.7} & 18.5~\gain{4.1} & 29.5~\gain{6.9} & 46.2~\gain{14.9} & 32.7~\gain{2.5} & 13.5~\gain{3.3} & 30.8~\gain{6.9} \\
(d4)~$+$~\method{}-4B ($\mathrm{M0}$) & \textbf{50.7}~\gain{22.7} & \textbf{29.6}~\gain{4.3} & \textbf{21.6}~\gain{7.2} & \textbf{34.0}~\gain{11.4} & \textbf{53.6}~\gain{22.3} & \textbf{35.0}~\gain{4.8} & \textbf{15.2}~\gain{5.0} & \textbf{34.6}~\gain{10.7} \\
\midrule
(e1)~\textit{Mixed first stage} & 9.7~\nodelta & 25.3~\nodelta & 17.6~\nodelta & 17.5~\nodelta & 10.3~\nodelta & 30.2~\nodelta & 11.8~\nodelta & 17.4~\nodelta \\
(e2)~$+$~Qwen3-Reranker-4B & 21.3~\gain{11.6} & 27.0~\gain{1.7} & 19.9~\gain{2.3} & 22.7~\gain{5.2} & 19.1~\gain{8.8} & 32.7~\gain{2.5} & 14.3~\gain{2.5} & 22.0~\gain{4.6} \\
(e3)~$+$~\method{}-4B ($\mathrm{M0}$) & \textbf{23.3}~\gain{13.6} & \textbf{29.6}~\gain{4.3} & \textbf{22.1}~\gain{4.5} & \textbf{25.0}~\gain{7.5} & \textbf{20.6}~\gain{10.3} & \textbf{35.0}~\gain{4.8} & \textbf{15.2}~\gain{3.4} & \textbf{23.6}~\gain{6.2} \\
\bottomrule
\end{tabular}%
}
\end{table}

\subsection{One-shot retrieval}\label{sec:oneshot-results}

\Cref{tab:oneshot-main} compares rerankers across four first-stage retrievers and three datasets in one-shot retrieval: each original question is searched once, without agent-generated follow-up queries.
Results combine BCP's two qrel sets and BRIGHT's and TRQA's subsets, with equal weight across datasets.
\Cref{app:oneshot} provides detailed results and standard deviations.

\method{} achieves the highest nDCG@10 and Recall@10 across all retriever--dataset combinations.
Dataset-averaged nDCG@10 gains are 9.8--12.8 pp over no reranking and 1.7--4.5 pp over Qwen3-Reranker-4B.
Both models use the same context-free prompt, so the gains reflect fine-tuning rather than additional inference context.
Improvements over no reranking are largest on BCP (13.6--23.0 pp), versus 4.3--8.8 pp on BRIGHT and TRQA.
Jina-Reranker-v3 improves BM25 but reduces BRIGHT Recall@10 for all three dense retrievers, by up to 4.0 pp.

For the subsequent agentic experiments, we use one retriever per dataset to limit cost: BM25 on BCP, \mbox{Nemotron-3-Embed-8B} on BRIGHT, and \mbox{AgentIR-4B} on TRQA (group~e).
For each dataset, we choose the retriever with which \method{} provides the smallest one-shot nDCG@10 improvement over no reranking.
The agentic evaluation therefore uses the least favorable first-stage settings for \method{} under this one-shot criterion.
We omit Jina-Reranker-v3, which consistently trails both other rerankers, and compare \method{} with no reranking and Qwen3-Reranker-4B throughout the remaining experiments.

\FloatBarrier
\begin{figure}[!t]
\centering
\includegraphics[width=\textwidth]{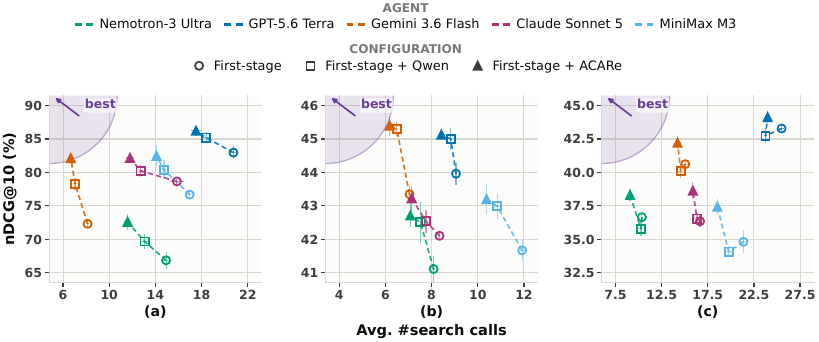}
\caption{\textbf{Agentic retrieval quality versus search effort}: upper left is better.
Panels: (a) BCP, (b) BRIGHT, (c) TRQA; axis ranges differ.
Dashed lines connect each agent's reranking configurations.
Qwen denotes Qwen3-Reranker-4B.
Results: \Cref{app:agentic-benchmarks}.}
\label{fig:ndcg-vs-calls}
\end{figure}

\subsection{Agentic retrieval}\label{sec:agentic-results}

\begin{table}[!tp]
\centering
\setlength{\tabcolsep}{5pt}
\small
\caption{\textbf{Agentic retrieval}, averaged equally over BCP, BRIGHT, and TRQA, using BM25, \mbox{Nemotron-3-Embed-8B}, and \mbox{AgentIR-4B}, respectively.
Each group starts without reranking, followed by the two rerankers.
Per-benchmark results: \Cref{app:agentic-benchmarks}.}
\label{tab:agentic-main}
\fitappendixtable{%
\begin{tabular}{@{}l lllll@{}}
\toprule
\textbf{Agent LLM} & \textbf{nDCG@10}$\uparrow$ & \textbf{Recall@10}$\uparrow$ & \multicolumn{3}{c}{\textbf{Avg. per query stats}} \\
\cmidrule(lr){4-6}
$+$\textbf{Reranker} & & & \textbf{Searches}$\downarrow$ & \textbf{LLM turns}$\downarrow$ & \textbf{Cost (\textcent)}$\downarrow$ \\
\midrule
(a1)~Nemotron-3 Ultra & 48.2~\std{0.4} & 46.3~\std{0.3} & 11.1~\std{0.1} & 12.2~\std{0.1} & 18.8~\std{0.2} \\
(a2)~$+$~Qwen3-Reranker-4B & 49.3~\std{0.5}$^{\dagger}$ & 48.1~\std{0.6}$^{\dagger}$ & 10.3~\std{0.1}$^{\dagger}$ & 11.3~\std{0.1}$^{\dagger}$ & 17.1~\std{0.6}$^{\dagger}$ \\
(a3)~$+$~\method{}-4B ($\mathrm{M3}$) & \textbf{51.2}~\std{0.6}$^{\dagger*}$ & \textbf{49.5}~\std{0.5}$^{\dagger*}$ & \textbf{9.2}~\std{0.1}$^{\dagger*}$ & \textbf{10.3}~\std{0.1}$^{\dagger*}$ & \textbf{15.0}~\std{0.6}$^{\dagger*}$ \\
\midrule
(b1)~MiniMax M3 & 51.1~\std{0.3} & 50.2~\std{0.2} & 16.8~\std{0.1} & 8.8~\std{0.1} & 6.5~\std{0.5} \\
(b2)~$+$~Qwen3-Reranker-4B & 52.5~\std{0.5}$^{\dagger}$ & 52.1~\std{0.4}$^{\dagger}$ & 15.1~\std{0.2}$^{\dagger}$ & 8.0~\std{0.1}$^{\dagger}$ & 5.9~\std{0.4} \\
(b3)~$+$~\method{}-4B ($\mathrm{M3}$) & \textbf{54.4}~\std{0.6}$^{\dagger*}$ & \textbf{53.4}~\std{0.6}$^{\dagger*}$ & \textbf{14.3}~\std{0.1}$^{\dagger*}$ & \textbf{7.7}~\std{0.1}$^{\dagger*}$ & \textbf{5.8}~\std{0.2} \\
\midrule
(c1)~Gemini 3.6 Flash & 52.1~\std{0.3} & 49.5~\std{0.2} & 10.1~\std{0.1} & 11.1~\std{0.1} & 13.9~\std{0.6} \\
(c2)~$+$~Qwen3-Reranker-4B & 54.6~\std{0.2}$^{\dagger}$ & 52.6~\std{0.3}$^{\dagger}$ & 9.4~\std{0.1}$^{\dagger}$ & 10.4~\std{0.1}$^{\dagger}$ & \textbf{12.9}~\std{0.5} \\
(c3)~$+$~\method{}-4B ($\mathrm{M2}$) & \textbf{56.6}~\std{0.1}$^{\dagger*}$ & \textbf{54.5}~\std{0.2}$^{\dagger*}$ & \textbf{9.0}~\std{0.1}$^{\dagger*}$ & \textbf{10.0}~\std{0.1}$^{\dagger*}$ & 13.2~\std{0.5} \\
\midrule
(d1)~GPT-5.6 Terra & 56.7~\std{0.4} & 54.2~\std{0.4} & 18.4~\std{0.2} & 4.2~\std{0.0} & 28.2~\std{0.6} \\
(d2)~$+$~Qwen3-Reranker-4B & 57.6~\std{0.2}$^{\dagger}$ & 54.9~\std{0.1}$^{\dagger}$ & 17.0~\std{0.1}$^{\dagger}$ & 4.0~\std{0.0}$^{\dagger}$ & \textbf{25.0}~\std{0.2}$^{\dagger}$ \\
(d3)~$+$~\method{}-4B ($\mathrm{M2}$) & \textbf{58.5}~\std{0.4}$^{\dagger*}$ & \textbf{55.9}~\std{0.3}$^{\dagger*}$ & \textbf{16.6}~\std{0.1}$^{\dagger*}$ & \textbf{3.9}~\std{0.0}$^{\dagger*}$ & 25.2~\std{0.6}$^{\dagger}$ \\
\midrule
(e1)~Claude Sonnet 5 & 52.4~\std{0.2} & 51.5~\std{0.2} & 13.6~\std{0.2} & 7.3~\std{0.1} & 36.5~\std{0.9} \\
(e2)~$+$~Qwen3-Reranker-4B & 53.1~\std{0.2}$^{\dagger}$ & 52.0~\std{0.2}$^{\dagger}$ & 12.3~\std{0.1}$^{\dagger}$ & 6.6~\std{0.1}$^{\dagger}$ & 31.3~\std{0.3}$^{\dagger}$ \\
(e3)~$+$~\method{}-4B ($\mathrm{M2}$) & \textbf{54.7}~\std{0.2}$^{\dagger*}$ & \textbf{53.4}~\std{0.2}$^{\dagger*}$ & \textbf{11.6}~\std{0.2}$^{\dagger*}$ & \textbf{6.3}~\std{0.1}$^{\dagger*}$ & \textbf{30.3}~\std{0.8}$^{\dagger}$ \\
\bottomrule
\end{tabular}%
}
\end{table}

\paragraph{Retrieval quality and search effort.}
Using BM25 on BCP, \mbox{Nemotron-3-Embed-8B} on BRIGHT, and \mbox{AgentIR-4B} on TRQA, \method{} improves retrieval quality while reducing search effort across all five agents.
Over Qwen3-Reranker-4B, dataset-averaged nDCG@10 and Recall@10 rise by 0.9--2.0 pp each, with 0.3--1.0 fewer searches per query (2.0--10.0\%; \Cref{tab:agentic-main}).
Quality gains and LLM-turn reductions are statistically significant for all five agents (\Cref{app:stats}).
Against no reranking, \method{} gains 1.8--4.5 nDCG@10 pp and saves 1.0--2.4 searches per query.

\Cref{fig:ndcg-vs-calls} shows higher mean nDCG@10 than the base reranker in all 15 agent--dataset combinations and fewer searches in 14.
Eight retain significant gains in at least one quality metric after correction.
On BCP and BRIGHT, searches fall by 0.3--1.5 per query (4.2--11.3\%) for every agent, with nine of ten reductions statistically significant.
\mbox{Nemotron-3-Embed-8B}'s training includes synthetic data generated from the BRIGHT corpus, which may leave less room for agentic search and reranking to improve quality.
Replacing it with the general-purpose \mbox{Qwen3-Embedding-4B} increases \method{}'s gain over the base reranker from 0.2 to 0.9 nDCG@10 pp for Nemotron-3 Ultra (\Cref{app:headroom}).

\looseness=-1 On TRQA, adding the context-blind base reranker on top of \mbox{AgentIR-4B} lowers mean nDCG@10 for four agents, whereas \method{} improves all five.
Context-blind reranking can thus erode a context-aware retriever's ranking quality; \method{} instead builds on it.
Over the base reranker, nDCG@10 gains are significant for all five agents and Recall@10 gains for four.
Searches fall for four agents, significantly for Ultra and M3; Terra's small increase is nonsignificant.
Across all 15 agent--dataset combinations, \method{} improves mean nDCG@10 by 0.9--9.8 pp and reduces searches by 4.7--25.6\% over no reranking.

\paragraph{Agent API cost.}
Both rerankers reduce mean agent API cost for every agent relative to no reranking: by 5.1--20.1\% with \method{} and 7.6--14.2\% with Qwen3-Reranker-4B.
\method{} additionally saves 12\% over Qwen3-Reranker-4B for Nemotron-3 Ultra, a statistically significant reduction.
For the other four agents, the difference between rerankers is not statistically significant across five runs.
Repeated document tokens and variable implicit caching can partially obscure savings from fewer searches (\Cref{app:cost}).
\Cref{sec:depth-latency} examines whether fewer searches also shorten end-to-end runs.

\section{Analysis}\label{sec:ablations}

\subsection{Latency and reranking depth}\label{sec:depth-latency}

\begin{figure}[!htbp]
\centering
\includegraphics[width=0.868\textwidth]{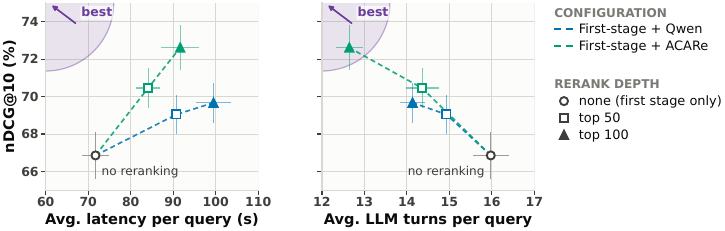}
\caption{\textbf{Reranking depth on BCP} with Nemotron-3 Ultra and BM25.
nDCG@10 versus latency (left) and LLM turns (right), averaged per query; upper left is better.
Values: \Cref{tab:ablation-depth}.}
\label{fig:depth-latency}
\end{figure}

To measure the latency benefit of reduced search effort, we compare reranking depths of 50 and 100 with Nemotron-3 Ultra on BCP using BM25 (\Cref{fig:depth-latency}).
Against Qwen3-Reranker-4B, \method{} reduces mean end-to-end latency by 7--8\% while improving nDCG@10.
The latency reduction at depth 50 and quality gains at depth 100 are statistically significant.
Both rerankers remain slower than no reranking.
\Cref{app:depth-latency} gives the configuration and detailed results.

\subsection{Effect of inference context}\label{sec:context-ablation}

We evaluate the same \method{} checkpoint in modes $\mathrm{M0}$--$\mathrm{M3}$ on BCP with Nemotron-3 Ultra and BM25 (\Cref{tab:ablation-context}).
In $\mathrm{M0}$, \method{} achieves 70.8 nDCG@10, compared with 69.7 for Qwen3-Reranker-4B.
The mean rises to 71.9 in $\mathrm{M1}$, 72.1 in $\mathrm{M2}$, and 72.6 in $\mathrm{M3}$.
Compared with $\mathrm{M0}$, $\mathrm{M3}$ reduces searches from 12.7 to 11.6 per query and achieves the highest mean Recall@10 and lowest mean API cost among the four modes.
In $\mathrm{M3}$, all five metrics improve statistically significantly over Qwen3-Reranker-4B and no reranking.
These results assess inference-time context for one checkpoint, not training with versus without context.

\begin{table}[!tp]
\centering
\setlength{\tabcolsep}{4pt}
\small
\caption{\textbf{Inference context} on BCP with Nemotron-3 Ultra and BM25.
Tests use this study's correction family (\Cref{app:stats}); marks may differ for shared runs in \Cref{tab:agentic-bcp,tab:agentic-bright,tab:agentic-trqa}.}
\label{tab:ablation-context}
\fitappendixtable{%
\begin{tabular}{@{}l lllll@{}}
\toprule
\textbf{Reranker, context} & \textbf{nDCG@10}$\uparrow$ & \textbf{Recall@10}$\uparrow$ & \multicolumn{3}{c}{\textbf{Avg. per query stats}} \\
\cmidrule(lr){4-6}
 & & & \textbf{Searches}$\downarrow$ & \textbf{LLM turns}$\downarrow$ & \textbf{Cost (\textcent)}$\downarrow$ \\
\midrule
No reranker & 66.9~\std{1.3} & 68.6~\std{1.0} & 14.9~\std{0.4} & 16.0~\std{0.4} & 35.0~\std{1.8} \\
Qwen3-Reranker-4B & 69.7~\std{1.1}$^{\dagger}$ & 72.2~\std{1.5}$^{\dagger}$ & 13.1~\std{0.3}$^{\dagger}$ & 14.1~\std{0.3}$^{\dagger}$ & 30.8~\std{0.8}$^{\dagger}$ \\
\midrule
\method{}-4B ($\mathrm{M0}$) & 70.8~\std{0.9}$^{\dagger}$ & 73.4~\std{1.3}$^{\dagger}$ & 12.7~\std{0.4}$^{\dagger}$ & 13.8~\std{0.4}$^{\dagger}$ & 33.0~\std{2.1} \\
\method{}-4B ($\mathrm{M1}$) & 71.9~\std{0.6}$^{\dagger*}$ & 74.9~\std{0.6}$^{\dagger}$ & 12.2~\std{0.3}$^{\dagger*}$ & 13.2~\std{0.3}$^{\dagger*}$ & 31.9~\std{1.3} \\
\method{}-4B ($\mathrm{M2}$) & 72.1~\std{1.6}$^{\dagger}$ & 74.8~\std{1.8}$^{\dagger}$ & 12.0~\std{0.3}$^{\dagger*}$ & 13.0~\std{0.3}$^{\dagger*}$ & 31.3~\std{1.2}$^{\dagger}$ \\
\method{}-4B ($\mathrm{M3}$) & \textbf{72.6}~\std{1.2}$^{\dagger*}$ & \textbf{75.2}~\std{0.9}$^{\dagger*}$ & \textbf{11.6}~\std{0.3}$^{\dagger*}$ & \textbf{12.6}~\std{0.3}$^{\dagger*}$ & \textbf{25.9}~\std{1.4}$^{\dagger*}$ \\
\bottomrule
\end{tabular}%
}
\end{table}

\FloatBarrier

\subsection{Small vs. large agents}\label{sec:size-cost}

\begin{wrapfigure}{r}{0.48\textwidth}
\vspace{-20pt}
\centering
\includegraphics[width=\linewidth]{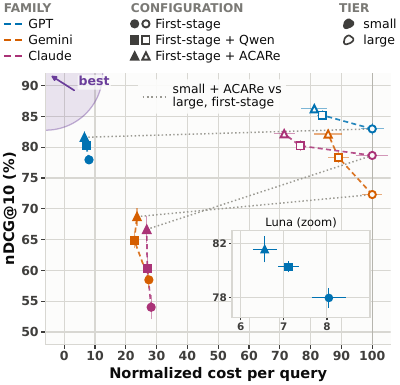}
\caption{\raggedright \textbf{Small vs.\ large agents on BCP} with BM25.
Pairs (small/large): \mbox{GPT-5.6} Luna/Terra; Gemini 3.5 Flash-Lite/3.6 Flash; Claude Haiku 4.5/Sonnet 5.
Qwen: Qwen3-Reranker-4B; \method{}: \method{}-4B.
Cost is relative to the large agent without reranking (100); up and left is better.
Inset: Luna zoom.
See \mbox{\Cref{tab:ablation-size}}.}
\label{fig:size-cost}
\end{wrapfigure}

We compare three pairs of smaller and larger agents on BCP with BM25: GPT-5.6 Luna and Terra, Gemini 3.5 Flash-Lite and 3.6 Flash, and Claude Haiku 4.5 and Sonnet 5.
Small and large denote the compared model tiers; exact parameter counts are not disclosed.
\Cref{fig:size-cost} compares quality and cost.

Relative to their larger counterparts without reranking, Luna, Flash-Lite, and Haiku close 73\%, 74\%, and 51\% of the nDCG@10 gap with \method{}, versus 46\%, 46\%, and 26\% with Qwen3-Reranker-4B.
When both agents in each pair use \method{}, the smaller agents incur 8\%, 28\%, and 38\% of their larger counterparts' API cost, with nDCG@10 scores 4.6, 13.4, and 15.5 pp lower, respectively (\Cref{app:size-cost}).

All six agents improve mean nDCG@10 over Qwen3-Reranker-4B with \method{}.
Haiku gains 6.3 pp, while Flash-Lite and Flash each gain approximately 3.8 pp.
Both rerankers lower mean API cost for every agent; \Cref{app:size-cost} gives detailed results and significance tests.

\par\vspace{1.5\baselineskip}
\WFclear
\FloatBarrier
\subsection{Transfer to end-to-end answer generation}\label{sec:piika}

We evaluate Piika on 57 BrowseComp-Plus$_{\mathrm{CM}}$ questions using BM25 API retrieval over ClimbMix's 553 million documents~\citep{diao2025nemotron}.
GPT-5.6 Luna and Nemotron-3.5 Lightning use no reranking, Qwen3-Reranker-4B, or \method{}-4B.
\method{} uses $\mathrm{M3}$ for Lightning and $\mathrm{M2}$ for Luna, which exposes no reasoning trace.
We omit monetary cost: Lightning switched between API providers and local serving under heavy throttling; Luna's separate vLLM reranker servers had different loads, so queuing delays changed agent-side prompt-cache reuse (\Cref{app:piika}).

\begin{table}[!tp]
\centering
\setlength{\tabcolsep}{4pt}
\caption{\textbf{Piika answer generation} on BrowseComp-Plus$_{\mathrm{CM}}$ (BM25 API).}
\label{tab:piika}
\vspace{2pt}
\fitappendixtable{%
\begin{tabular}{@{}lllll@{}}
\toprule
\textbf{Agent LLM} & \textbf{Answer accuracy} & \textbf{Recall@all} & \textbf{Precision@all} & \textbf{Searches} \\
$+$\textbf{Reranker} & (\%) $\uparrow$ & (\%) $\uparrow$ & (\%) $\uparrow$ & / query $\downarrow$ \\
\midrule
(a1)~GPT-5.6 Luna & 62.9~\std{2.8} & 9.3~\std{0.6} & 20.4~\std{2.0} & 28.0~\std{2.1} \\
(a2)~$+$~Qwen3-Reranker-4B & \textbf{71.1}~\std{2.4} & 9.3~\std{0.4} & 21.3~\std{1.1} & 24.8~\std{2.6} \\
(a3)~$+$~ACARe-4B ($\mathrm{M2}$) & \textbf{71.1}~\std{2.1}$^{\dagger}$ & \textbf{9.9}~\std{0.9} & \textbf{24.8}~\std{0.9}$^{\dagger*}$ & \textbf{22.8}~\std{1.0}$^{\dagger}$ \\
\midrule
(b1)~Nemotron-3.5 Lightning & 10.2~\std{0.7} & 1.4~\std{0.5} & 5.8~\std{0.6} & 11.1~\std{0.8} \\
(b2)~$+$~Qwen3-Reranker-4B & 18.7~\std{5.0} & 1.6~\std{0.2} & 6.8~\std{0.4} & 11.5~\std{2.1} \\
(b3)~$+$~ACARe-4B ($\mathrm{M3}$) & \textbf{19.3}~\std{3.5}$^{\dagger}$ & \textbf{2.1}~\std{0.6} & \textbf{9.4}~\std{1.7}$^{\dagger}$ & \textbf{10.8}~\std{0.8} \\
\bottomrule
\end{tabular}}
\end{table}

Across six runs (\Cref{tab:piika}), \method{} significantly improves accuracy and precision over no reranking for both agents, with the fewest mean searches.
With Luna, it also significantly reduces searches and matches Qwen3's mean accuracy with significantly higher precision.
Lightning's mean accuracy nearly doubles (10.2\% to 19.3\%); its limited capacity, about 3B active parameters, may partly explain its low accuracy and recall (\Cref{app:piika}).

We assess context and fine-tuning with Lightning on a separate stratified sample of 200 BCP questions, using six runs per configuration (\Cref{tab:piika-context,app:piika-context}).
With Qwen3-Reranker-4B's weights fixed, mean accuracy rises from 48.2\% in $\mathrm{M0}$ to 54.8\% in $\mathrm{M2}$ and 55.9\% in $\mathrm{M3}$; \method{} in $\mathrm{M3}$ reaches 57.6\%.
Recall and precision average gold and evidence qrel scores and rise as searches fall at each step.
\method{} significantly improves all four metrics over Qwen3 in $\mathrm{M0}$ ($p\leq0.002$), and recall, precision, and searches over Qwen3 in $\mathrm{M2}$.
With identical $\mathrm{M3}$ inputs, \method{} leads on all four means, adding 1.7 pp accuracy and 1.0 pp precision; these final gains are not significant after correction.

\begin{inplacetable}
\centering
\setlength{\tabcolsep}{3pt}
\caption{\textbf{BCP context control} with Nemotron-3.5 Lightning and BM25 in Piika.
$*$: significant difference from sub-query-only Qwen3-Reranker-4B.}
\label{tab:piika-context}
\fitappendixtable{%
\begin{tabular}{@{}llllll@{}}
\toprule
\textbf{Reranker} & \textbf{Input} & \textbf{Accuracy} $\uparrow$ & \textbf{Recall@all} $\uparrow$ & \textbf{Precision@all} $\uparrow$ & \textbf{Searches} $\downarrow$ \\
\midrule
No reranker & -- & 42.2~\std{1.6} & 44.3~\std{1.3} & 11.8~\std{0.5} & 10.8~\std{0.8} \\
Qwen3-Reranker-4B & $q_t$ ($\mathrm{M0}$) & 48.2~\std{2.3} & 54.2~\std{1.4} & 14.4~\std{0.6} & 10.1~\std{0.6} \\
Qwen3-Reranker-4B & $q_t,Q$ ($\mathrm{M2}$) & 54.8~\std{2.3} & 61.2~\std{1.2} & 19.2~\std{0.6} & 9.3~\std{0.5} \\
Qwen3-Reranker-4B & $q_t,\tau_t,Q$ ($\mathrm{M3}$) & 55.9~\std{0.7} & 63.9~\std{0.9} & 20.1~\std{0.8} & 8.7~\std{0.4} \\
\method{}-4B ($\mathrm{M3}$) & $q_t,\tau_t,Q$ ($\mathrm{M3}$) & \textbf{57.6}~\std{1.8}$^{\dagger*}$ & \textbf{65.1}~\std{1.4}$^{\dagger*}$ & \textbf{21.2}~\std{0.5}$^{\dagger*}$ & \textbf{8.4}~\std{0.3}$^{\dagger*}$ \\
\bottomrule
\end{tabular}}
\end{inplacetable}

\section{Conclusion}\label{sec:conclusion}

Search-agent sub-queries can omit the goal and reasoning needed to judge document relevance.
We introduce \method{}, to our knowledge the first pointwise LLM reranker trained to use this context for search-agent sub-queries.
Fine-tuned from Qwen3-Reranker-4B, \method{} improves mean retrieval quality across three benchmarks and five LLM families while reducing search effort.
On TRQA, it improves context-aware AgentIR's results for all five agents, whereas Qwen3-Reranker-4B lowers mean nDCG@10 for four.
On BCP, smaller agents gain at least as much as larger ones over no reranking, closing 51--74\% of the nDCG@10 gap to larger agents without reranking, versus 26--46\% with the base.
The benefits extend to answer generation through a BM25 API over 553 million documents.
A BCP context control shows that \method{}'s context formulation also benefits the unchanged base reranker, while the trained checkpoint achieves the best means on every metric.
These results highlight \method{} as a promising lightweight reranker whose independent, parallelizable scoring makes deeper reranking practical to explore.

\section{Limitations and future work}\label{sec:limitations}

We expand the training data through an agent rollout with a second retriever, while retaining the original DR-Synth seed questions, agent LLM, and oracle reranker.
Varying these components, the corpus, and the retrievers and rerankers used for negative mining could broaden supervision and further improve reranking effectiveness.

We evaluate on two existing deep-search datasets with fixed corpora, BCP and TRQA.
The base-reranker context control uses one agent on BCP.
Evaluation on additional such datasets as they become available and with live-web search APIs remains future work.

Our cost analysis covers agent API usage, assuming agent LLMs are substantially larger and costlier than retrievers and rerankers.
In our size study, smaller agents benefit more from a higher-quality retrieval tool, but further reducing their size toward the 4B reranker could make reranking costs non-negligible.
Measuring total serving costs would clarify this effectiveness--cost tradeoff.

\section*{Ethics statement}

\method{} is fine-tuned from Qwen3-Reranker-4B and may inherit biases and other limitations of the base model.
Training examples are mined using an oracle reranker, whose judgments may introduce biases into the training dataset.
Fine-tuning for relevance does not ensure that highly ranked documents are accurate, fair, or safe.
Downstream applications should account for these risks when using the retrieved evidence.

\section*{AI-use statement}

We used Claude and OpenAI Codex to draft and polish some of the manuscript sections, identify and examine related literature, assist with code development, and check analyses.
LLMs also generated synthetic training trajectories, provided relevance supervision, and judged answers, as described in \Cref{app:training,app:piika}.
The authors thoroughly reviewed and vetted every section of the manuscript through multiple rounds of revision, including AI-assisted text, claims, citations, and analyses.
The authors retain full responsibility for the final manuscript and all reported results.

\section*{Reproducibility statement}

\paragraph{Data generation and training.}
\Cref{sec:method} and \Cref{app:training} specify the data sources, seed questions and corpus, rollout agents and retrievers, oracle reranker and answer judge, and prompts used to construct supervision.
We describe re-mining, filtering, positive and negative selection, and the size of each component of the final training pool.
The paper also provides the four context templates, context-dropout procedure, exact scoring and loss functions, optimization settings, sequence limits, training hardware, and the validation protocol used to select hyperparameters and the data mixture.
Together, these details document the construction of the training data and the fine-tuning procedure.

\paragraph{Evaluations and analyses.}
\Cref{sec:setup,sec:ablations} and \Cref{app:repro} specify benchmark versions and evaluated samples, relevance labels, retriever and reranker configurations, agent models and reasoning settings, and search and output depths.
We document harness behavior, tool interfaces, context limits, stopping and recovery rules, and prompt sources; the Piika agent and judge prompts are reproduced from retained execution logs.
Detailed per-dataset and analysis results appear in \begingroup\renewcommand{\creflastconjunction}{, and~}\Cref{app:oneshot,app:agentic-benchmarks,app:depth-latency,app:size-cost,app:headroom,app:piika}\endgroup.
\Cref{app:cost} explains token-based cost accounting and the included charges.
All reported evaluations aggregate at least five runs; \Cref{app:stats} defines the aggregation, variability estimates, statistical tests, and correction families, and distinguishes tested from descriptive comparisons.
Exact API responses may vary with generation randomness and model updates.

This is a work in progress.
We will release the trained checkpoint, training data, and all model-training and evaluation code soon to support independent reproduction and extension of this work.

\FloatBarrier
\bibliography{iclr2027_conference}
\bibliographystyle{iclr2027_conference}

\clearpage
\appendix
\makeatletter
\setlength{\@fptop}{0pt}
\setlength{\@fpsep}{8pt}
\makeatother

\section{Training details}\label{app:training}
\suppressfloats[t]

We fine-tune \method{} using NeMo Automodel\footnote{\url{https://github.com/NVIDIA-NeMo/Automodel}} with FSDP2 and activation checkpointing on eight A100 80\,GB GPUs.
\Cref{tab:training} summarizes the training configuration.
All training-input truncation removes excess tokens from the end, retaining the beginning of each field.
\Cref{fig:prompt-modes} gives the prompt templates for the context modes defined in \Cref{sec:context}.

\paragraph{Model development and final training.}
During model development, we held out 10\% of the training pool as a validation set, assigning all instances of each original question to a single side of the split.
We selected hyperparameters and the training-data mixture using LCE loss, accuracy (the positive ranked first within its group), and mean reciprocal rank (MRR) on the validation split.
With the recipe fixed, we trained \method{}-4B once on the full pool without a validation holdout.

\begin{inplacetable}
\centering
\setlength{\tabcolsep}{4pt}
\small
\caption{\textbf{Training configuration} for \method{}-4B.
The data pool is described in \Cref{app:training}; the context modes are defined in \Cref{sec:context}, and their prompts are shown in \Cref{fig:prompt-modes}.}
\label{tab:training}
\fitappendixtable{%
\begin{tabular}{@{}l l@{}}
\toprule
\textbf{Setting} & \textbf{Value} \\
\midrule
\multicolumn{2}{@{}l}{\textit{Model}} \\
\quad Initialization & Qwen3-Reranker-4B~\citep{zhang2025qwen3embed} \\
\quad Fine-tuning & full, bf16; no adapters and no new parameters \\
\quad Scoring head & pretrained LM head; score $\sigma(\ell_{\mathrm{yes}}-\ell_{\mathrm{no}})$ \\
\addlinespace[2pt]
\multicolumn{2}{@{}l}{\textit{Data}} \\
\quad Training pool & 16,601 instances over 54{,}838 distinct passages \\
\quad Labelings & three over two rollouts (5,104 $+$ 5,219 $+$ 6,278) \\
\quad Group & 1 positive $+$ 7 retrieved negatives, mined for the same turn \\
\quad In-batch negatives & none \\
\quad Context dropout & $\tau_t$ and $Q$ dropped independently, $p=0.5$ each \\
\addlinespace[2pt]
\multicolumn{2}{@{}l}{\textit{Objective}} \\
\quad Loss & localized contrastive; cross-entropy over the group, positive at index 0 \\
\quad Temperature & 0.5, applied to logit differences \\
\addlinespace[2pt]
\multicolumn{2}{@{}l}{\textit{Sequence}} \\
\quad Maximum length & 5{,}120 tokens \\
\quad Per-field caps & $Q$ 2{,}048, $\tau_t$ 512, $q_t$ 256; the document takes the remainder \\
\addlinespace[2pt]
\multicolumn{2}{@{}l}{\textit{Optimization}} \\
\quad Framework & NeMo Automodel \\
\quad Distributed training & FSDP2 with activation checkpointing \\
\quad Optimizer & AdamW, weight decay 0.01 \\
\quad Learning rate & $5\times10^{-6}$, 20 warm-up steps, then cosine decay \\
\quad Gradient clipping & global norm 1.0 \\
\quad Batch size & 32 global; 1 group per device step, gradient accumulation 4 \\
\quad Epochs & 1, i.e.\ about 519 optimizer steps \\
\quad Seed & 42, for both data order and context dropout \\
\addlinespace[2pt]
\multicolumn{2}{@{}l}{\textit{Hardware}} \\
\quad GPUs & 8$\times$ A100 80\,GB \\
\quad Wall clock & about 10 hours \\
\bottomrule
\end{tabular}%
}
\end{inplacetable}

\begin{figure}[!htb]
\begin{mdframed}[font=\footnotesize, roundcorner=10pt, linewidth=1pt, innerleftmargin=10pt, innerrightmargin=10pt, innertopmargin=5pt, innerbottommargin=5pt]
\textbf{System message} (identical in all four modes)
\\ \\
Judge whether the Document meets the requirements based on the Query and the Instruct provided. Note that the answer can only be ``yes'' or ``no''.
\\ \\
\rule{\linewidth}{0.4pt}
\\
\textbf{M0 --- no context}
\\
$\langle$Instruct$\rangle$: Given a web search query, retrieve relevant passages that answer the query\\
$\langle$Query$\rangle$: \{$sub\_query$\}\\
$\langle$Document$\rangle$: \{$document$\}
\\ \\
\rule{\linewidth}{0.4pt}
\\
\textbf{M1 --- reasoning trace}
\\
$\langle$Instruct$\rangle$: Given a user generated web search query and the user's reasoning trace that motivated the query (when available), retrieve relevant passages that answer the query in light of the reasoning context\\
$\langle$Query$\rangle$: \#Reasoning Trace: \{$reasoning\_trace$\}\\
\hspace*{1em}\#Query: \{$sub\_query$\}\\
$\langle$Document$\rangle$: \{$document$\}
\\ \\
\rule{\linewidth}{0.4pt}
\\
\textbf{M2 --- original question}
\\
$\langle$Instruct$\rangle$: Given a user generated query issued during an iterative research session and the original question (global query) that motivated the session, retrieve relevant passages that answer the query in light of the original question\\
$\langle$Query$\rangle$: \#Original Question (global query): \{$global\_query$\}\\
\hspace*{1em}\#Query: \{$sub\_query$\}\\
$\langle$Document$\rangle$: \{$document$\}
\\ \\
\rule{\linewidth}{0.4pt}
\\
\textbf{M3 --- reasoning trace and original question}
\\
$\langle$Instruct$\rangle$: Given a user generated query, the original question (global query), and the user's reasoning trace (when available) that motivated the query, retrieve relevant passages that answer the query in light of the original question and reasoning context\\
$\langle$Query$\rangle$: \#Original Question (global query): \{$global\_query$\}\\
\hspace*{1em}\#Reasoning Trace: \{$reasoning\_trace$\}\\
\hspace*{1em}\#Query: \{$sub\_query$\}\\
$\langle$Document$\rangle$: \{$document$\}
\end{mdframed}
\caption{The four context modes \method{} is trained to score in.
Only the instruction and the contents of the query block change; the system message and the document field are the same throughout.}
\label{fig:prompt-modes}
\end{figure}

\paragraph{Rollouts and supervision.}
We use 500 WebShaper questions~\citep{tao2025webshaper} with gold answers and supporting documents.
Both rollouts use the AgentIR harness\footnote{\url{https://github.com/texttron/AgentIR}} over the same approximately 1.1M-document corpus.
The rollout agent is \mbox{Tongyi-DeepResearch-30B-A3B}; GLM-4.6 is the oracle reranker.
The original DR-Synth rollout uses \mbox{Qwen3-Embedding-8B}~\citep{chen2026agentir}, and our new rollout uses \mbox{Nemotron-3-Embed-8B}.
Evaluation uses neither the rollout agent nor its oracle; Retrieval Bench and Piika also differ from the rollout harness.

The oracle ranks up to 50 candidates per sub-query, including prepended gold supporting documents, using the original question and gold answer.
Its top document supplies the positive and its seven lowest-ranked documents supply distinct negatives; its top five documents are returned to the rollout agent.
Nemotron-3 Ultra judges final-answer correctness.
With the same harness, agent, and oracle, the new rollout has 276 correct trajectories out of 500 (55.2\%), compared with 249 (49.8\%) in the original DR-Synth rollout.
This is an observed difference between the two rollouts, without repeated-rollout estimates of variability.
Only successful trajectories contribute training instances.
\Cref{fig:rollout-prompt-agent,fig:rollout-prompt-oracle,fig:rollout-prompt-judge} reproduce the AgentIR rollout, oracle, and judge prompts~\citep{chen2026agentir}.

\paragraph{Filtering and training pool.}
We exclude New York Times passages, at the publisher's request, before constructing the final pool.
For the original DR-Synth labels, removing rows with an excluded passage in any position reduces 5{,}238 instances to 5{,}104.
For re-mined labels, we remove rows with an excluded positive and replace excluded negatives with other ranked candidates.
This removes 19 re-mined rows, leaving 5{,}219; thus the original and re-mined labels retain different numbers of instances from the same 249 trajectories.

Together with 6{,}278 instances from the 276 new trajectories, the pool contains 16{,}601 instances over 54{,}838 distinct passages.
Every instance retains its original question.
In the re-mined component, 15.7\% of instances have malformed reasoning traces that we treat as missing; these instances remain usable in modes without a trace.

\paragraph{Scoring and objective.}
Let $x_t$ denote the prompt containing the sub-query $q_t$ and the retained context fields.
For a candidate document $d$, define the logit difference and inference score as
\begin{equation}
 z_\theta(x_t,d)=\ell_\theta(\text{yes}\mid x_t,d)-\ell_\theta(\text{no}\mid x_t,d),
 \qquad s_\theta(x_t,d)=\sigma\!\left(z_\theta(x_t,d)\right).
 \label{eq:reranker-score}
\end{equation}
Here $\ell_\theta$ is the next-token logit at the final non-padding position.
For a training group $D_t=(d_t^+,d_{t,1}^-,\ldots,d_{t,m}^-)$, localized contrastive estimation (LCE)~\citep{10.1007/978-3-030-72240-1_26} minimizes
\begin{equation}
 \mathcal{L}_{\mathrm{LCE}}(x_t,D_t)
 =-\log\frac{\exp\!\left(z_\theta(x_t,d_t^+)/T\right)}
 {\exp\!\left(z_\theta(x_t,d_t^+)/T\right)
 +\sum_{j=1}^{m}\exp\!\left(z_\theta(x_t,d_{t,j}^-)/T\right)},
 \qquad T=0.5.
 \label{eq:lce}
\end{equation}
The loss uses the raw logit differences before sigmoid and is averaged over training groups.
Each group contains one positive at index 0 and seven distinct negatives mined for the same sub-query; no in-batch negatives are used.
The temperature applies only to the training objective.
The sigmoid preserves document ordering at inference.

\suppressfloats[t]

\begin{figure}[!htb]
\begin{mdframed}[font=\footnotesize, roundcorner=10pt, linewidth=1pt, innerleftmargin=10pt, innerrightmargin=10pt, innertopmargin=5pt, innerbottommargin=5pt]
\setlength{\parindent}{0pt}\setlength{\parskip}{0.3\baselineskip}%
You are a deep research assistant. Your core function is to conduct thorough, multi-source investigations into any topic. You must handle both broad, open-domain inquiries and queries within specialized academic fields. For every request, synthesize information from credible, diverse sources to deliver a comprehensive, accurate, and objective response. When you have gathered sufficient information and are ready to provide the definitive response, you must enclose the entire final answer within \textless{}answer\textgreater{}\textless{}/answer\textgreater{} tags.

\rule{\linewidth}{0.4pt}\par
\textbf{\# Tools}

You may call the search tool one or multiple times to assist with the user query.

You are provided with the search tool, its signature within \textless{}tools\textgreater{}\textless{}/tools\textgreater{} XML tags:\\
\textless{}tools\textgreater{}\\
\{"type": "function", "function": \{"name": "search", "description": "Perform Google web searches then returns a string of the top search results. Accepts a single query.", "parameters": \{"type": "object", "properties": \{"query": \{"type": "string", "description": "The search query."\}\}, "required": ["query"]\}\}\}\\
\textless{}/tools\textgreater{}

For each search tool call, return a json object with function name and arguments within \textless{}tool\_call\textgreater{}\textless{}/tool\_call\textgreater{} XML tags:\\
\textless{}tool\_call\textgreater{}\\
\{"name": "search", "arguments": \textless{}args-json-object\textgreater{}\}\\
\textless{}/tool\_call\textgreater{}

Current date:
\end{mdframed}
\caption{Rollout-agent prompt from the AgentIR release~\citep{chen2026agentir}.
Its template names Google web search, but the handler calls the configured corpus retriever.
AgentIR appends the execution date in YYYY-MM-DD format to the last line.}
\label{fig:rollout-prompt-agent}
\end{figure}

\begin{figure}[!htb]
\begin{mdframed}[font=\footnotesize, roundcorner=10pt, linewidth=1pt, innerleftmargin=10pt, innerrightmargin=10pt, innertopmargin=5pt, innerbottommargin=5pt]
\setlength{\parindent}{0pt}\setlength{\parskip}{0.3\baselineskip}%
\textbf{System message}\\
You are an intelligent assistant that can rank passages to best help an user in correctly answering a question. You will be given a complex question, and its correct answer. A user, who is trying to solve the complex question, issues a simpler probing web query, which retrieved some passages. You are to rank these passages based on their relevancy to the user's probing query, while prioritizing their usefulness in helping the user to correctly answer the complex question.

\rule{\linewidth}{0.4pt}\par
\textbf{Prefix}\\
A user is trying to answer a complex question: \{question\}. The correct answer is: \{correct\_answer\}. A user, trying to solve the complex question, issued this probing web query: \{query\}, which retrieved \{num\} passages. I will provide you with these \{num\} passages, each indicated by a numerical identifier []. Rank the passages based on their relevance to the probing query, as well as their usefulness in helping the user to correctly answer the complex question.

\rule{\linewidth}{0.4pt}\par
\textbf{Body (one line per candidate)}\\{}
[\{rank\}] \{candidate\}

\rule{\linewidth}{0.4pt}\par
\textbf{Suffix}\\
Complex Question: \{question\}.\\
Correct Answer: \{correct\_answer\}.\\
Probing Query: \{query\}.\\
Rank the \{num\} passages above based on their relevance to the probing query, as well as their usefulness in helping the user to correctly answer the complex question. All the passages should be included and listed using identifiers, in descending order of relevance. The output format should be [] \textgreater{} [], e.g., [2] \textgreater{} [1], Answer concisely and directly and only respond with the ranking results, do not say any word or explain.
\end{mdframed}
\caption{Oracle listwise prompt from the AgentIR release~\citep{chen2026agentir}, assembled from a system message, a prefix, one body line per candidate, and a suffix.
Both the agent and oracle receive the seed question; only the oracle receives its gold answer.
\{num\} is at most 50 in our mining.}
\label{fig:rollout-prompt-oracle}
\end{figure}

\begin{figure}[!htb]
\begin{mdframed}[font=\footnotesize, roundcorner=10pt, linewidth=1pt, innerleftmargin=10pt, innerrightmargin=10pt, innertopmargin=5pt, innerbottommargin=5pt]
\setlength{\parindent}{0pt}\setlength{\parskip}{0.3\baselineskip}%
Judge whether the following [response] to [question] is correct or not based on the precise and unambiguous [correct\_answer] below.

[question]: \{question\}

[response]: \{response\}

[correct\_answer]: \{correct\_answer\}

Your judgment must be in the format and criteria specified below:

extracted\_final\_answer: The final exact answer extracted from the [response]. 

[correct\_answer]: Repeat the [correct\_answer] given above.

reasoning: Explain why the extracted\_final\_answer is correct or incorrect based on [correct\_answer], in the context of this [question]. You should judge whether the extracted\_final\_answer is semantically equivalent to [correct\_answer], allowing the extracted\_final\_answer to be string variations of [correct\_answer]. You should also allow the extracted\_final\_answer to be more precise or verbose than [correct\_answer], as long as its additional details are correct. Do not comment on any background to the problem, do not attempt to solve the problem, do not argue for any answer different than [correct\_answer], focus only on whether the answers are semantically equivalent.

correct: Answer 'yes' if extracted\_final\_answer matches the [correct\_answer] given above, or is within a small margin of error for numerical problems. Answer 'no' otherwise, i.e. if there is any inconsistency, ambiguity, non-equivalency, or if the extracted answer is incorrect.

confidence: The extracted confidence score between 0\% and 100\% from [response]. Put 100 if there is no confidence score available.
\end{mdframed}
\caption{Judge prompt adapted from the AgentIR release~\citep{chen2026agentir}, with spelling and typographic corrections.
It decides whether a trajectory may contribute to the training pool.
Only its \texttt{correct} field is used: all turns from trajectories judged \texttt{yes} enter the training pool; trajectories judged \texttt{no} are excluded.}
\label{fig:rollout-prompt-judge}
\end{figure}

\FloatBarrier

\section{Experimental setup details}\label{app:repro}

\Cref{tab:eval} reports benchmark sizes and shared evaluation settings.
For one-shot retrieval and Retrieval Bench, samples are drawn with seed 42 and capped at 200 questions per subset.
Each BCP run is evaluated against both gold and evidence qrels; BRIGHT includes all 1{,}384 questions; TRQA contributes 569 questions across its three subsets.
We will release the selection seed and the complete list of selected queries with the evaluation code so the evaluation samples can be reproduced exactly.

\paragraph{Benchmarks.}
BCP~\citep{chen-etal-2026-browsecomp} evaluates multi-hop questions over a fixed corpus of 100{,}195 documents.
For its 830 questions, evidence qrels identify documents needed to answer the question, while gold qrels identify those that also contain the final answer.
Each run produces one ranking, which we score against both qrel sets and average; these are not separate experiments.
Following the original BCP paper's setup, we retain the first 512 tokens of each document for evaluation, limiting both reranker input and the document text retained in the agent conversation.

BRIGHT~\citep{su2024bright} contains 1{,}384 questions in twelve domains, each with its own corpus.
Corpus sizes range from 7{,}894 documents for Pony to 413{,}932 for LeetCode.
Relevance often requires reasoning beyond lexical matching.
The number of relevant documents also varies by domain, from about two per question in the theorem and code subsets to twenty in Pony.

TRQA~\citep{rafiee2026trqa} contains 2{,}327 test questions across Wiki1, Wiki2, and synthetic e-commerce.
Each requires a single numerical answer aggregating relevant entities, such as a sum, average, or count, so missing evidence can change the answer.
We use the later released \texttt{wiki\_partial} corpus (5.9M passages) for both Wikipedia subsets, alongside the 3.3M-passage e-commerce corpus.\footnote{\url{https://huggingface.co/datasets/mahtaa/trqa}}
Unlike BCP, BRIGHT and TRQA have no separate 512-token document cap; model sequence limits still apply.

\begin{inplacetable}
\centering
\setlength{\tabcolsep}{5pt}
\small
\caption{\textbf{Evaluation configuration.}
Corpus size is the number of documents; BRIGHT gives a range because each domain has its own corpus.
\textbf{Sampled} is the number of queries actually evaluated, and \textbf{Retriever} the fixed first stage used in the agentic experiments.
BCP uses two qrel sets (gold and evidence) to score the same run.
Only BCP applies separate document truncation.}
\label{tab:eval}
\fitappendixtable{%
\begin{tabular}{@{}l lll@{}}
\toprule
& \begin{tabular}[b]{@{}c@{}}\textbf{BrowseComp-Plus}\\\textbf{\citep{chen-etal-2026-browsecomp}}\end{tabular} & \begin{tabular}[b]{@{}c@{}}\textbf{BRIGHT}\\\textbf{\citep{su2024bright}}\end{tabular} & \begin{tabular}[b]{@{}c@{}}\textbf{Total Recall QA}\\\textbf{\citep{rafiee2026trqa}}\end{tabular} \\
\midrule
Corpus size & 100{,}195 & 7{,}894--413{,}932 & 3.3M--5.9M \\
Questions & 830 & 1{,}384 & 2{,}327 \\
Sampled & 200 & 1{,}384 & 569 \\
Subsets & 1 & 12 domains & 3 (2 Wikipedia, 1 e-commerce) \\
Retriever & BM25 & Nemotron-3-Embed-8B & AgentIR-4B \\
Doc.\ truncation & 512 tokens & none & none \\
\midrule
\multicolumn{4}{@{}l}{\textit{Shared across benchmarks}} \\
\quad Retrieval depth & \multicolumn{3}{@{}l}{top 100 candidates per call} \\
\quad Rerank depth & \multicolumn{3}{@{}l}{all 100 (\Cref{sec:depth-latency} is the only experiment that varies this)} \\
\quad Model input cap & \multicolumn{3}{@{}l}{8{,}192 tokens, retriever and reranker alike} \\
\quad Documents per search & \multicolumn{3}{@{}l}{$k=10$ by default, unseen documents only} \\
\quad Query sample & \multicolumn{3}{@{}l}{drawn with seed 42, capped at 200 queries per subset} \\
\quad Runs per configuration & \multicolumn{3}{@{}l}{5; tables report mean $\pm$ sample standard deviation} \\
\quad Significance test & \multicolumn{3}{@{}l}{See \Cref{app:stats}} \\
\quad Maximum trajectory length & \multicolumn{3}{@{}l}{100 turns} \\
\quad Concurrency & \multicolumn{3}{@{}l}{2 queries per run} \\
\bottomrule
\end{tabular}%
}
\end{inplacetable}

\paragraph{Retrieval and reranking.}
On BRIGHT and TRQA, BM25 uses Pyserini's default parameter values, $k_1=0.9$ and $b=0.4$~\citep{Lin_etal_SIGIR2021_Pyserini}.
On BCP, both one-shot and agentic experiments deliberately use $k_1=16$ and $b=1$, the long-document setting identified by the grid search of \citet{hsu2026piserini}, rather than the defaults.
In one-shot retrieval, \mbox{AgentIR-4B} receives the original question and the literal string \texttt{Empty} for absent reasoning.
In agentic retrieval, it also receives the current reasoning trace when available.
Other retrievers use only the query.

\paragraph{Inference input limits.}
Before prompt assembly, \method{} caps the original question at 2{,}048 tokens, an available reasoning trace at 512 tokens, and the sub-query at 256 tokens in both evaluation harnesses.
All input truncation removes excess tokens from the end, retaining the beginning of each field or document.
Dense-retriever inputs and the combined inputs to \method{} and Qwen3-Reranker-4B are capped at 8{,}192 tokens.

\paragraph{Retrieval Bench.}\label{app:agentic} Each trajectory starts with ten documents retrieved for the original question and reranked when reranking is enabled.
The agent then follows a ReAct-style loop~\citep{yao2023react}, using \texttt{search} (dense) or \texttt{keyword\_search} (BM25), which replace the upstream \texttt{retrieve} tool.
\Cref{fig:retrieval-bench-system_a,fig:retrieval-bench-system_b,fig:retrieval-bench-tools} reproduce the public upstream system prompt and tool descriptions.

\begin{figure}[!t]
\begin{mdframed}[font=\footnotesize, roundcorner=10pt, linewidth=1pt, innerleftmargin=10pt, innerrightmargin=10pt, innertopmargin=5pt, innerbottommargin=5pt]
\setlength{\parindent}{0pt}\setlength{\parskip}{0.5\baselineskip}
You are a retrieval agent that finds all documents related to a given query.

<Goal>\\
You are given a search query and a list of documents retrieved for that query. Your task is to write new queries and use the given search tool to find *ALL* the related and somewhat related documents to the given query (i.e., maximize recall).\\
If the user's query is a question, you should not answer the question yourself. Instead, you should find the related documents for the given query.\\
</Goal>

<RELEVANCE\_DEFINITION>\\
- You should be careful, in the context of this task, what it means to be a "query", "document", and "relevant" can sometimes be very complex and might not follow the traditional definition of these terms in standard information retrieval.\\
- In standard retrieval, a query is usually a user question (like a web search query), the document is some sort of content that provides information (e.g., a web page), and these two are considered relevant if the document provides information that answers the user's query.\\
- However, in our setting, this could be different. Here are some examples:\\
    * the query is a programming problem and documents are programming language syntax references. A document is relevant if it contains the reference for the programming syntax used for solving the problem.\\
    * both query and documents are descriptions programming problems and a query and document are relevant if the same approach is used to solve them.\\
    * the query is a math problem and documents are theorems. Relevant documents (theorems) are the ones that are useful for solving the math problem.\\
    * the query and document are both math problems. A query and a document are relevant if the same theorem is used for solving them.\\
    * the query is a task description (e.g., for an API programmer) and documents are descriptions of available APIs. Relevant documents (e.g., APIs) are the ones needed for completing the task.\\
- This is not an exhaustive list. These are just some examples to show you the complexity of queries, documents, and the concept of relevance in this task.\\
- Note that even here, the relevant documents are still the ones that are useful for a user who is searching for the given query. But the relation is more nuanced.\\
- You should analyze the query and some of the available documents. And then reason about what could be a meaningful definition of relevance in this case, and what the user could be looking for.\\
- Moreover, sometimes, the query could be even a prompt that is given to a Large Language Model (LLM) and the user wants to find the useful documents for the LLM that help answering/solving this prompt.\\
</RELEVANCE\_DEFINITION>
\end{mdframed}
\caption{Retrieval Bench system prompt (part 1): task and extended relevance definition.}
\label{fig:retrieval-bench-system_a}
\end{figure}

A search retrieves and, when enabled, reranks 100 candidates unless the experiment varies the depth.
It returns the highest-ranked $k$ unseen documents, with $k=10$ by default.
Previously seen documents are listed by identifier rather than repeating their content in the tool response.
Agents can change $k$: across the evaluated agents and benchmarks, searches return 9.4--10.3 new documents on average, except for GPT-5.6 Terra on BCP, which averages 11.8 new documents per search.

\begin{figure}[!tp]
\begin{mdframed}[font=\footnotesize, roundcorner=10pt, linewidth=1pt, innerleftmargin=10pt, innerrightmargin=10pt, innertopmargin=5pt, innerbottommargin=5pt]
\setlength{\parindent}{0pt}\setlength{\parskip}{0.3\baselineskip}
<WORKFLOW>\\
- You are given a retrieval tool, powered by a dense embedding model, that takes a text query and returns the most similar documents.\\
- As explained above, reason and figure out what the meaning of relevance is in this case, and what could be relevant and useful information for the given query.\\
- You can call the search tool multiple times.\\
- Search for related documents to the user's query from different angles.\\
- If needed, revise your search queries based on the documents you find in previous steps.\\
- Once you are confident that you have found all the related and somewhat related documents and there are no more related documents in the corpus, call the "final\_results" tool to finish the task.\\
- When calling "final\_results", you must select exactly the 10 most relevant documents among all documents you have retrieved.\\
- When calling the "final\_results" tool, the list of documents must be sorted in the decreasing level of relevance to the query. I.e., the first document is the most relevant to the query, the second document is the second most relevant to the query, and so on.\\
</WORKFLOW>

<BEST\_PRACTICES>\\
- You should be thorough and find all related and somewhat related documents.\\
- The goal is to increase the **Recall** of your search attempt. So, if multiple documents are relevant to the given query, you should find and report all of them even if only a subset of them is enough for answering the query.\\
- **TIP**: you can look at the list of documents retrieved using the original query and think what other queries you can use to find the potentially related documents that are missing in these results.\\
</BEST\_PRACTICES>
\end{mdframed}
\caption{Retrieval Bench system prompt (part 2): workflow and search guidance, with initial documents enabled and a final list of ten documents.}
\label{fig:retrieval-bench-system_b}
\end{figure}

\begin{table}[!tp]
\centering
\setlength{\tabcolsep}{4pt}
\caption{\textbf{Agent configuration, token prices, and token-priced costs.}
Prices are USD per million tokens; -- denotes no supplied cache rate or charge.
Totals are computed before rounding.
Totals count retained runs once, including both Piika studies and their answer judging.
Lightning includes local-inference estimates; see \Cref{app:cost}.}
\label{tab:agents}\label{tab:pricing}
\fitappendixtable{%
\begin{tabular*}{\linewidth}{@{\extracolsep{\fill}}lllll@{}}
\toprule
\multicolumn{5}{@{}l}{\textit{Agent configuration}} \\
\textbf{Agent LLM} & \textbf{Provider} & \textbf{Model identifier} & \textbf{Reasoning} & \textbf{Context} \\
\midrule
Nemotron-3 Ultra & NIM & \texttt{nvidia/nemotron-3-ultra-nvfp4} & high & 262K \\
\addlinespace[2pt]
MiniMax M3 & NIM & \texttt{minimaxai/minimax-m3} & high & 512K \\
\addlinespace[2pt]
Gemini 3.6 Flash & NIM & \texttt{google/gemini-3.6-flash} & high & 1M \\
\addlinespace[2pt]
GPT-5.6 Terra & NIM & \texttt{openai/gpt-5.6-terra} & high & 1M \\
\addlinespace[2pt]
Claude Sonnet 5 & NIM & \texttt{anthropic/claude-sonnet-5} & adaptive & 1M \\
\addlinespace[2pt]
GPT-5.6 Luna & \begin{tabular}[t]{@{}l@{}}NIM (size study);\\OpenAI (Piika)\end{tabular} & \texttt{openai/gpt-5.6-luna} & high & 1M \\
\addlinespace[2pt]
Gemini 3.5 Flash-Lite & NIM & \texttt{google/gemini-3.5-flash-lite} & high & 1M \\
\addlinespace[2pt]
Claude Haiku 4.5 & NIM & \texttt{anthropic/claude-haiku-4-5} & fixed budget & 200K \\
\addlinespace[2pt]
Nemotron-3.5 Lightning & \begin{tabular}[t]{@{}l@{}}OpenRouter;\\local serving\end{tabular} & \texttt{nvidia/nemotron-3.5-lightning} & high & 262K \\
\bottomrule
\end{tabular*}}
\par\medskip
\fitappendixtable{%
\begin{tabular*}{\linewidth}{@{\extracolsep{\fill}}llllll@{}}
\toprule
\textbf{LLM} & \multicolumn{4}{c}{\textbf{USD per million tokens}} & \textbf{Token-priced} \\
 & \textbf{Input} & \textbf{Cache read} & \textbf{Cache write} & \textbf{Output} & \textbf{cost (USD)} \\
\midrule
Nemotron-3 Ultra & 0.6 & 0.12 & -- & 2.4 & 7{,}495.87 \\
MiniMax M3 & 0.3 & 0.06 & -- & 1.2 & 1{,}511.12 \\
Gemini 3.6 Flash & 0.75 & 0.075 & -- & 3.75 & 3{,}732.40 \\
GPT-5.6 Terra & 2.00 & 0.2 & 2.5 & 12.00 & 5{,}861.11 \\
Claude Sonnet 5 & 2.00 & 0.2 & 2.5 & 10.00 & 7{,}533.48 \\
GPT-5.6 Luna & 0.2 & 0.02 & 0.25 & 1.2 & 152.69 \\
Gemini 3.5 Flash-Lite & 0.3 & 0.03 & -- & 2.5 & 101.21 \\
Claude Haiku 4.5 & 1.00 & 0.1 & 1.25 & 5.00 & 523.35 \\
Nemotron-3.5 Lightning & 0.08 & 0.04 & -- & 0.2 & 123.92 \\
\midrule
GPT-5.5 (judge; OpenAI) & 5.00 & 0.5 & -- & 30.00 & 64.21 \\
\midrule
\multicolumn{5}{@{}l}{Total} & 27{,}099.35 \\
\bottomrule
\end{tabular*}}
\end{table}

A turn may contain multiple search calls.
Nemotron-3 Ultra and Gemini 3.6 Flash average approximately one per turn, MiniMax M3 and Claude Sonnet 5 approximately two, and GPT-5.6 Terra more than four.
The trajectory is limited to 100 turns.
The search agent selects and ranks documents from its accumulated evidence, then submits their IDs through \texttt{final\_results}.
This tool records the supplied ranking and terminates the search.
We compute nDCG@10 and Recall@10 on the first ten submitted IDs, or on the recovered ranking if the search agent fails.
\label{app:recovery}Across the five agents, three reranking configurations, and five runs on all benchmarks, 0.145\% of query executions fail before recovery.
The most common failures are output-token limits, followed by failed LLM calls and turn limits.
On failure, the harness invokes a separate selection agent using the same LLM to review all documents retrieved throughout the search and select a ranked subset.
If selection fails, the harness uses reciprocal rank fusion (RRF) over the accumulated retrieval rankings.
After recovery, no query execution is left without a final ranking.

\paragraph{Agent serving and sampling.}
\Cref{tab:agents} lists agent routes, reasoning controls, context windows, token prices, and token-priced usage totals for the reported experiments.
Retrieval Bench agents, including Luna in the size study, are accessed through an NVIDIA NIM gateway using LiteLLM.\footnote{\url{https://docs.litellm.ai/}}
In Piika, Luna uses OpenAI directly, while Lightning uses OpenRouter and local serving; the GPT-5.5 answer judge also uses OpenAI directly.
The completion limit for Claude Sonnet 5 is 16{,}384 tokens because the default of 4{,}096 truncated its responses.
Other agents retain their default completion limits; \Cref{tab:agents} lists each agent's reasoning setting.
Each Retrieval Bench run processes two questions concurrently.

\FloatBarrier

\begin{figure}[!t]
\begin{mdframed}[font=\footnotesize, roundcorner=10pt, linewidth=1pt, innerleftmargin=10pt, innerrightmargin=10pt, innertopmargin=5pt, innerbottommargin=5pt]
\setlength{\parindent}{0pt}\setlength{\parskip}{0.5\baselineskip}
\textbf{retrieve}

Search for documents related to a query using dense retrieval.

\textit{Parameters (source descriptions):}

\textbf{query} (string, required): The search query.

\textbf{top\_k} (integer, optional, default 10): Number of documents to retrieve.

\textbf{final\_results}

Signals the completion of the search process for the current query.

Use this tool when:\\
- You have found all the relevant documents to the query.\\
- Despite several attempts, you cannot find good documents for the given query.

The message should include:\\
- A brief summary of your exploration and the results\\
- Explanation if the search was unsuccessful

When reporting the selected document IDs, make sure:\\
- the list of document IDs is sorted in the decreasing level of relevance to the query. I.e., the first document in the list is the most relevant to the query, the second is the second most relevant to the query, and so on.\\
- You must choose exactly 10 document IDs when calling this function.

The successful\_search field should be set to true if you believed you have found the most relevant documents to the user's query, and false otherwise. And partial if it is in between.

\textit{Parameters (source descriptions):}

\textbf{message} (string, required): A message for the user to explain why you think you found all the related documents and there is no related document is missing. Also, include a short description of your exploration process. If your attempts to find related documents were unsuccessful, explain why.

\textbf{doc\_ids} (array of string, required): List of document IDs that are relevant to the user's query sorted descending by their level of relevance to the user's query. I.e., the first document is the most relevant to the query, the second is the second most relevant to the query, and so on.

\textbf{search\_successful} (string, required, values true, false, partial): Whether you managed to find all the related documents to the query.

\end{mdframed}
\caption{Retrieval Bench retrieval and final-ranking tool descriptions and parameter schemas.
Each search retrieves 100 candidates and returns ten unseen documents by default; the agent can change the return count.
The final ranking contains ten documents.}
\label{fig:retrieval-bench-tools}
\end{figure}

\FloatBarrier

\section{Per-dataset one-shot results}\label{app:oneshot}

\Cref{tab:oneshot-bcp,tab:oneshot-bright,tab:oneshot-trqa} give the one-shot results by subset, using the row identifiers from \Cref{tab:oneshot-main}.
BCP's average combines gold and evidence labels; BRIGHT and TRQA averages weight their twelve and three subsets equally.
The main table averages benchmarks equally before rounding.

No-reranker and Jina-Reranker-v3 results are identical across all five runs.
The two 4B rerankers have small run-to-run variation, with standard deviations of at most 0.7 pp. A displayed standard deviation of 0.0 can represent a value below 0.05 pp.

\begin{table}[!tp]
\centering
\setlength{\tabcolsep}{3pt}
\caption{\textbf{One-shot retrieval on BCP.}
Avg.
combines the gold and evidence qrel sets.}
\label{tab:oneshot-bcp}
\fitappendixtable{%
\begin{tabular}{@{}l lll lll@{}}
\toprule
\textbf{Retriever} & \multicolumn{3}{c}{\textbf{nDCG@10}} & \multicolumn{3}{c}{\textbf{Recall@10}} \\
\cmidrule(lr){2-4}\cmidrule(lr){5-7}
$+$\textbf{Reranker} & Gold & Evidence & Avg. & Gold & Evidence & Avg. \\
\midrule
(a1)~BM25 & 8.9~\std{0.0} & 10.4~\std{0.0} & 9.7~\std{0.0} & 10.3~\std{0.0} & 10.3~\std{0.0} & 10.3~\std{0.0} \\
(a2)~$+$~Jina-Reranker-v3 & 16.3~\std{0.0} & 18.0~\std{0.0} & 17.2~\std{0.0} & 17.1~\std{0.0} & 15.8~\std{0.0} & 16.4~\std{0.0} \\
(a3)~$+$~Qwen3-Reranker-4B & 20.1~\std{0.0} & 22.4~\std{0.0} & 21.3~\std{0.0} & 19.9~\std{0.0} & 18.3~\std{0.0} & 19.1~\std{0.0} \\
(a4)~$+$~\method{}-4B ($\mathrm{M0}$) & \textbf{21.6}~\std{0.1} & \textbf{25.0}~\std{0.0} & \textbf{23.3}~\std{0.1} & \textbf{20.8}~\std{0.0} & \textbf{20.3}~\std{0.0} & \textbf{20.6}~\std{0.0} \\
\midrule
(b1)~Qwen3-Embedding-4B & 12.8~\std{0.0} & 13.1~\std{0.0} & 12.9~\std{0.0} & 15.9~\std{0.0} & 13.5~\std{0.0} & 14.7~\std{0.0} \\
(b2)~$+$~Jina-Reranker-v3 & 23.7~\std{0.0} & 23.6~\std{0.0} & 23.6~\std{0.0} & 29.4~\std{0.0} & 22.8~\std{0.0} & 26.1~\std{0.0} \\
(b3)~$+$~Qwen3-Reranker-4B & 31.5~\std{0.0} & 31.0~\std{0.0} & 31.3~\std{0.0} & 34.4~\std{0.0} & 27.8~\std{0.0} & 31.1~\std{0.0} \\
(b4)~$+$~\method{}-4B ($\mathrm{M0}$) & \textbf{35.7}~\std{0.0} & \textbf{36.0}~\std{0.0} & \textbf{35.9}~\std{0.0} & \textbf{39.5}~\std{0.0} & \textbf{31.8}~\std{0.0} & \textbf{35.7}~\std{0.0} \\
\midrule
(c1)~AgentIR-4B & 25.0~\std{0.0} & 28.4~\std{0.0} & 26.7~\std{0.0} & 32.0~\std{0.0} & 28.5~\std{0.0} & 30.2~\std{0.0} \\
(c2)~$+$~Jina-Reranker-v3 & 31.8~\std{0.0} & 31.6~\std{0.0} & 31.7~\std{0.0} & 40.5~\std{0.0} & 30.9~\std{0.0} & 35.7~\std{0.0} \\
(c3)~$+$~Qwen3-Reranker-4B & 44.4~\std{0.0} & 44.7~\std{0.0} & 44.6~\std{0.0} & 50.3~\std{0.0} & 40.7~\std{0.0} & 45.5~\std{0.0} \\
(c4)~$+$~\method{}-4B ($\mathrm{M0}$) & \textbf{48.5}~\std{0.0} & \textbf{48.9}~\std{0.0} & \textbf{48.7}~\std{0.0} & \textbf{54.7}~\std{0.0} & \textbf{44.7}~\std{0.0} & \textbf{49.7}~\std{0.0} \\
\midrule
(d1)~Nemotron-3-Embed-8B & 28.0~\std{0.0} & 28.0~\std{0.0} & 28.0~\std{0.0} & 34.8~\std{0.0} & 27.7~\std{0.0} & 31.3~\std{0.0} \\
(d2)~$+$~Jina-Reranker-v3 & 31.0~\std{0.0} & 30.5~\std{0.0} & 30.8~\std{0.0} & 37.2~\std{0.0} & 29.2~\std{0.0} & 33.2~\std{0.0} \\
(d3)~$+$~Qwen3-Reranker-4B & 44.6~\std{0.0} & 41.6~\std{0.0} & 43.1~\std{0.0} & 53.7~\std{0.1} & 38.8~\std{0.0} & 46.2~\std{0.0} \\
(d4)~$+$~\method{}-4B ($\mathrm{M0}$) & \textbf{52.0}~\std{0.0} & \textbf{49.4}~\std{0.0} & \textbf{50.7}~\std{0.0} & \textbf{61.5}~\std{0.0} & \textbf{45.6}~\std{0.0} & \textbf{53.6}~\std{0.0} \\
\bottomrule
\end{tabular}%
}
\end{table}

\clearpage

\begin{sidewaystable}[p]
\begin{minipage}{0.94\textwidth}
\centering
\setlength{\tabcolsep}{2pt}
\renewcommand{\arraystretch}{0.78}
\caption{\textbf{One-shot retrieval on BRIGHT.}
Avg.
is the macro-average over twelve subsets.}
\label{tab:oneshot-bright}
\fitappendixtable{%
\begin{tabular}{@{}l@{\hspace{6pt}} rrrrrrrrrrrrr@{}}
\toprule
\shortstack[l]{\textbf{Retriever}\\$+$\textbf{Reranker}} & Bio. & Earth & Econ. & Psy. & Rob. & StackO. & Sust. & Leet. & Pony & AoPS & TQA-Q & TQA-T & Avg. \\
\midrule
\multicolumn{14}{@{}l}{\textit{nDCG@10}} \\
\midrule
(a1)~BM25 & 17.6~\std{0.0} & 28.7~\std{0.0} & 15.5~\std{0.0} & 14.4~\std{0.0} & 12.3~\std{0.0} & 17.8~\std{0.0} & 15.8~\std{0.0} & 26.1~\std{0.0} & 5.0~\std{0.0} & 6.1~\std{0.0} & 8.0~\std{0.0} & 1.7~\std{0.0} & 14.1~\std{0.0} \\
(a2)~$+$~Jina-Reranker-v3 & 22.8~\std{0.0} & 35.9~\std{0.0} & \textbf{21.8}~\std{0.0} & 29.1~\std{0.0} & 20.3~\std{0.0} & 18.6~\std{0.0} & 22.7~\std{0.0} & 20.4~\std{0.0} & \textbf{12.4}~\std{0.0} & 8.1~\std{0.0} & 10.8~\std{0.0} & \textbf{9.3}~\std{0.0} & 19.4~\std{0.0} \\
(a3)~$+$~Qwen3-Reranker-4B & 24.1~\std{0.1} & 37.5~\std{0.0} & 18.5~\std{0.0} & 28.6~\std{0.1} & 25.5~\std{0.0} & \textbf{28.6}~\std{0.0} & 22.1~\std{0.0} & 33.4~\std{0.0} & 10.1~\std{0.0} & 10.0~\std{0.0} & \textbf{12.4}~\std{0.0} & 8.9~\std{0.0} & 21.6~\std{0.0} \\
(a4)~$+$~\method{}-4B ($\mathrm{M0}$) & \textbf{27.4}~\std{0.0} & \textbf{41.6}~\std{0.1} & 21.5~\std{0.0} & \textbf{30.0}~\std{0.0} & \textbf{27.3}~\std{0.1} & 27.2~\std{0.1} & \textbf{23.8}~\std{0.1} & \textbf{33.6}~\std{0.0} & 11.0~\std{0.0} & \textbf{10.5}~\std{0.0} & \textbf{12.4}~\std{0.1} & 8.6~\std{0.0} & \textbf{22.9}~\std{0.0} \\
\midrule
(b1)~Qwen3-Embedding-4B & 17.0~\std{0.0} & 34.7~\std{0.0} & 16.5~\std{0.0} & 23.2~\std{0.0} & 11.7~\std{0.0} & 15.9~\std{0.0} & 16.7~\std{0.0} & 36.2~\std{0.0} & 1.2~\std{0.0} & \textbf{9.7}~\std{0.0} & 35.9~\std{0.0} & \textbf{41.3}~\std{0.0} & 21.7~\std{0.0} \\
(b2)~$+$~Jina-Reranker-v3 & 24.2~\std{0.0} & 37.9~\std{0.0} & 23.8~\std{0.0} & 27.3~\std{0.0} & 17.3~\std{0.0} & 15.6~\std{0.0} & 25.8~\std{0.0} & 19.1~\std{0.0} & 4.7~\std{0.0} & 5.3~\std{0.0} & 26.4~\std{0.0} & 34.0~\std{0.0} & 21.8~\std{0.0} \\
(b3)~$+$~Qwen3-Reranker-4B & 29.8~\std{0.0} & 37.1~\std{0.0} & 20.4~\std{0.0} & 26.7~\std{0.0} & 19.8~\std{0.0} & 26.5~\std{0.0} & 23.5~\std{0.2} & 39.8~\std{0.0} & \textbf{6.6}~\std{0.0} & 9.1~\std{0.0} & 39.2~\std{0.0} & 36.6~\std{0.1} & 26.3~\std{0.0} \\
(b4)~$+$~\method{}-4B ($\mathrm{M0}$) & \textbf{34.6}~\std{0.0} & \textbf{42.8}~\std{0.1} & \textbf{25.1}~\std{0.1} & \textbf{30.0}~\std{0.1} & \textbf{22.8}~\std{0.1} & \textbf{27.2}~\std{0.0} & \textbf{26.6}~\std{0.0} & \textbf{40.2}~\std{0.0} & 6.3~\std{0.0} & 8.6~\std{0.0} & \textbf{40.8}~\std{0.1} & 38.3~\std{0.1} & \textbf{28.6}~\std{0.0} \\
\midrule
(c1)~AgentIR-4B & 24.5~\std{0.0} & \textbf{43.9}~\std{0.0} & 17.9~\std{0.0} & 24.5~\std{0.0} & 14.2~\std{0.0} & 17.6~\std{0.0} & 13.8~\std{0.0} & 37.5~\std{0.0} & 2.8~\std{0.0} & 8.2~\std{0.0} & 36.0~\std{0.0} & 28.6~\std{0.0} & 22.5~\std{0.0} \\
(c2)~$+$~Jina-Reranker-v3 & 26.5~\std{0.0} & 37.4~\std{0.0} & 23.6~\std{0.0} & 29.3~\std{0.0} & 18.9~\std{0.0} & 16.9~\std{0.0} & 24.3~\std{0.0} & 18.8~\std{0.0} & 5.2~\std{0.0} & 5.5~\std{0.0} & 26.4~\std{0.0} & 32.2~\std{0.0} & 22.1~\std{0.0} \\
(c3)~$+$~Qwen3-Reranker-4B & 30.2~\std{0.1} & 37.8~\std{0.2} & 18.9~\std{0.0} & 27.6~\std{0.0} & 20.4~\std{0.0} & \textbf{28.9}~\std{0.0} & 23.0~\std{0.0} & 39.3~\std{0.0} & \textbf{14.5}~\std{0.1} & \textbf{9.4}~\std{0.0} & \textbf{40.4}~\std{0.1} & 37.1~\std{0.0} & 27.3~\std{0.0} \\
(c4)~$+$~\method{}-4B ($\mathrm{M0}$) & \textbf{36.7}~\std{0.1} & 43.4~\std{0.1} & \textbf{23.7}~\std{0.0} & \textbf{29.6}~\std{0.0} & \textbf{23.1}~\std{0.0} & 28.7~\std{0.0} & \textbf{26.1}~\std{0.1} & \textbf{40.6}~\std{0.0} & \textbf{14.5}~\std{0.0} & 9.1~\std{0.1} & \textbf{40.4}~\std{0.1} & \textbf{37.2}~\std{0.0} & \textbf{29.4}~\std{0.0} \\
\midrule
(d1)~Nemotron-3-Embed-8B & 33.8~\std{0.0} & 41.4~\std{0.0} & 23.8~\std{0.0} & 31.1~\std{0.0} & 20.1~\std{0.0} & 19.7~\std{0.0} & 22.3~\std{0.0} & 30.6~\std{0.0} & 3.5~\std{0.0} & \textbf{10.3}~\std{0.0} & 36.4~\std{0.0} & 30.1~\std{0.0} & 25.3~\std{0.0} \\
(d2)~$+$~Jina-Reranker-v3 & 25.2~\std{0.0} & 37.3~\std{0.0} & 23.3~\std{0.0} & 27.8~\std{0.0} & 20.4~\std{0.0} & 16.0~\std{0.0} & 25.2~\std{0.0} & 19.2~\std{0.0} & 6.5~\std{0.0} & 5.8~\std{0.0} & 27.5~\std{0.0} & 30.5~\std{0.0} & 22.1~\std{0.0} \\
(d3)~$+$~Qwen3-Reranker-4B & 30.5~\std{0.1} & 37.3~\std{0.0} & 21.8~\std{0.0} & 28.4~\std{0.0} & 22.1~\std{0.1} & 27.0~\std{0.0} & 22.9~\std{0.0} & 37.9~\std{0.0} & 13.8~\std{0.0} & 10.0~\std{0.0} & 35.8~\std{0.0} & 36.5~\std{0.3} & 27.0~\std{0.0} \\
(d4)~$+$~\method{}-4B ($\mathrm{M0}$) & \textbf{37.0}~\std{0.1} & \textbf{43.3}~\std{0.0} & \textbf{25.9}~\std{0.1} & \textbf{32.0}~\std{0.0} & \textbf{25.4}~\std{0.0} & \textbf{27.3}~\std{0.1} & \textbf{25.9}~\std{0.1} & \textbf{39.1}~\std{0.0} & \textbf{14.1}~\std{0.1} & 8.8~\std{0.0} & \textbf{37.7}~\std{0.0} & \textbf{38.7}~\std{0.1} & \textbf{29.6}~\std{0.0} \\
\midrule
\multicolumn{14}{@{}l}{\textit{Recall@10}} \\
\midrule
(a1)~BM25 & 20.7~\std{0.0} & 33.9~\std{0.0} & 15.1~\std{0.0} & 17.0~\std{0.0} & 16.3~\std{0.0} & 23.3~\std{0.0} & 20.1~\std{0.0} & 30.5~\std{0.0} & 2.7~\std{0.0} & 5.4~\std{0.0} & 9.4~\std{0.0} & 2.6~\std{0.0} & 16.4~\std{0.0} \\
(a2)~$+$~Jina-Reranker-v3 & 27.4~\std{0.0} & 39.6~\std{0.0} & 22.0~\std{0.0} & 33.5~\std{0.0} & 25.3~\std{0.0} & 25.1~\std{0.0} & \textbf{27.2}~\std{0.0} & 26.1~\std{0.0} & \textbf{6.5}~\std{0.0} & 8.1~\std{0.0} & 10.6~\std{0.0} & \textbf{9.2}~\std{0.0} & 21.7~\std{0.0} \\
(a3)~$+$~Qwen3-Reranker-4B & 27.9~\std{0.1} & 43.6~\std{0.0} & 24.3~\std{0.0} & 33.2~\std{0.0} & 29.5~\std{0.0} & \textbf{32.0}~\std{0.0} & 25.3~\std{0.0} & \textbf{39.6}~\std{0.0} & 4.6~\std{0.1} & 10.2~\std{0.0} & \textbf{12.8}~\std{0.1} & \textbf{9.2}~\std{0.0} & 24.4~\std{0.0} \\
(a4)~$+$~\method{}-4B ($\mathrm{M0}$) & \textbf{31.1}~\std{0.1} & \textbf{46.0}~\std{0.0} & \textbf{25.8}~\std{0.0} & \textbf{35.6}~\std{0.0} & \textbf{31.2}~\std{0.0} & \textbf{32.0}~\std{0.0} & 26.0~\std{0.2} & 39.0~\std{0.0} & 5.1~\std{0.0} & \textbf{10.7}~\std{0.1} & 12.4~\std{0.2} & \textbf{9.2}~\std{0.0} & \textbf{25.4}~\std{0.0} \\
\midrule
(b1)~Qwen3-Embedding-4B & 20.3~\std{0.0} & 40.8~\std{0.0} & 19.7~\std{0.0} & 29.3~\std{0.0} & 15.3~\std{0.0} & 22.9~\std{0.0} & 22.3~\std{0.0} & 43.1~\std{0.0} & 1.1~\std{0.0} & \textbf{10.2}~\std{0.0} & 39.0~\std{0.0} & \textbf{51.4}~\std{0.0} & 26.3~\std{0.0} \\
(b2)~$+$~Jina-Reranker-v3 & 28.7~\std{0.0} & 42.3~\std{0.0} & 26.2~\std{0.0} & 30.6~\std{0.0} & 19.7~\std{0.0} & 21.1~\std{0.0} & \textbf{31.6}~\std{0.0} & 21.9~\std{0.0} & 2.7~\std{0.0} & 6.7~\std{0.0} & 30.5~\std{0.0} & 44.0~\std{0.0} & 25.5~\std{0.0} \\
(b3)~$+$~Qwen3-Reranker-4B & 35.7~\std{0.1} & 44.4~\std{0.0} & 26.1~\std{0.0} & 33.4~\std{0.0} & 23.9~\std{0.0} & 34.6~\std{0.0} & 28.1~\std{0.0} & 53.4~\std{0.0} & \textbf{3.6}~\std{0.0} & 9.9~\std{0.0} & 45.3~\std{0.0} & 47.9~\std{0.7} & 32.2~\std{0.1} \\
(b4)~$+$~\method{}-4B ($\mathrm{M0}$) & \textbf{40.3}~\std{0.0} & \textbf{47.5}~\std{0.0} & \textbf{32.9}~\std{0.1} & \textbf{36.2}~\std{0.0} & \textbf{25.1}~\std{0.0} & \textbf{36.5}~\std{0.0} & 31.4~\std{0.0} & \textbf{53.5}~\std{0.0} & 3.5~\std{0.0} & 9.4~\std{0.0} & \textbf{46.2}~\std{0.1} & 50.7~\std{0.3} & \textbf{34.4}~\std{0.0} \\
\midrule
(c1)~AgentIR-4B & 29.7~\std{0.0} & 46.3~\std{0.0} & 20.3~\std{0.0} & 30.9~\std{0.0} & 19.6~\std{0.0} & 27.3~\std{0.0} & 20.4~\std{0.0} & 46.9~\std{0.0} & 2.1~\std{0.0} & 9.4~\std{0.0} & 41.5~\std{0.0} & 36.2~\std{0.0} & 27.6~\std{0.0} \\
(c2)~$+$~Jina-Reranker-v3 & 32.8~\std{0.0} & 41.1~\std{0.0} & 24.3~\std{0.0} & \textbf{34.2}~\std{0.0} & 22.7~\std{0.0} & 24.7~\std{0.0} & 29.3~\std{0.0} & 23.1~\std{0.0} & 3.6~\std{0.0} & 5.5~\std{0.0} & 31.1~\std{0.0} & 44.6~\std{0.0} & 26.4~\std{0.0} \\
(c3)~$+$~Qwen3-Reranker-4B & 36.4~\std{0.3} & 44.5~\std{0.1} & 24.5~\std{0.0} & 33.9~\std{0.0} & 25.4~\std{0.0} & 38.0~\std{0.0} & 29.1~\std{0.1} & \textbf{53.1}~\std{0.2} & 7.5~\std{0.0} & 9.9~\std{0.0} & 46.0~\std{0.1} & 47.9~\std{0.0} & 33.0~\std{0.0} \\
(c4)~$+$~\method{}-4B ($\mathrm{M0}$) & \textbf{44.1}~\std{0.1} & \textbf{48.7}~\std{0.2} & \textbf{31.5}~\std{0.0} & \textbf{34.2}~\std{0.0} & \textbf{25.6}~\std{0.0} & \textbf{39.5}~\std{0.0} & \textbf{31.2}~\std{0.2} & 51.8~\std{0.0} & \textbf{7.7}~\std{0.0} & \textbf{10.0}~\std{0.0} & \textbf{46.4}~\std{0.3} & \textbf{48.0}~\std{0.0} & \textbf{34.9}~\std{0.0} \\
\midrule
(d1)~Nemotron-3-Embed-8B & 39.1~\std{0.0} & 45.3~\std{0.0} & 27.2~\std{0.0} & 35.3~\std{0.0} & 23.9~\std{0.0} & 29.0~\std{0.0} & 29.5~\std{0.0} & 39.3~\std{0.0} & 2.2~\std{0.0} & 10.1~\std{0.0} & 39.8~\std{0.0} & 41.9~\std{0.0} & 30.2~\std{0.0} \\
(d2)~$+$~Jina-Reranker-v3 & 28.7~\std{0.0} & 40.7~\std{0.0} & 24.0~\std{0.0} & 32.9~\std{0.0} & 24.0~\std{0.0} & 21.5~\std{0.0} & \textbf{33.6}~\std{0.0} & 22.8~\std{0.0} & 4.0~\std{0.0} & 5.9~\std{0.0} & 32.6~\std{0.0} & 43.5~\std{0.0} & 26.2~\std{0.0} \\
(d3)~$+$~Qwen3-Reranker-4B & 37.5~\std{0.4} & 44.4~\std{0.0} & 28.9~\std{0.0} & 36.1~\std{0.0} & 26.7~\std{0.0} & 35.2~\std{0.0} & 27.6~\std{0.0} & \textbf{48.5}~\std{0.0} & 7.4~\std{0.0} & \textbf{11.3}~\std{0.1} & 40.9~\std{0.0} & \textbf{48.3}~\std{0.6} & 32.7~\std{0.1} \\
(d4)~$+$~\method{}-4B ($\mathrm{M0}$) & \textbf{43.8}~\std{0.2} & \textbf{48.3}~\std{0.1} & \textbf{34.9}~\std{0.0} & \textbf{39.7}~\std{0.0} & \textbf{28.8}~\std{0.0} & \textbf{36.5}~\std{0.1} & 30.9~\std{0.1} & 47.9~\std{0.1} & \textbf{7.5}~\std{0.0} & 9.9~\std{0.0} & \textbf{43.1}~\std{0.0} & \textbf{48.3}~\std{0.1} & \textbf{35.0}~\std{0.0} \\
\bottomrule
\end{tabular}%
}
\end{minipage}
\end{sidewaystable}

\clearpage

\begin{table}[!tp]
\centering
\setlength{\tabcolsep}{2.95pt}
\caption{\textbf{One-shot retrieval on TRQA.}
Avg.
is the macro-average over three subsets.}
\label{tab:oneshot-trqa}
\fitappendixtable{%
\begin{tabular}{@{}l llll llll@{}}
\toprule
\textbf{Retriever} & \multicolumn{4}{c}{\textbf{nDCG@10}} & \multicolumn{4}{c}{\textbf{Recall@10}} \\
\cmidrule(lr){2-5}\cmidrule(lr){6-9}
$+$\textbf{Reranker} & Wiki1 & Wiki2 & E-comm. & Avg. & Wiki1 & Wiki2 & E-comm. & Avg. \\
\midrule
(a1)~BM25 & 20.8\,\std{0.0} & 8.0\,\std{0.0} & 8.3\,\std{0.0} & 12.4\,\std{0.0} & 15.9\,\std{0.0} & 5.5\,\std{0.0} & 1.8\,\std{0.0} & 7.7\,\std{0.0} \\
(a2)~$+$~Jina-Reranker-v3 & 27.4\,\std{0.0} & 11.6\,\std{0.0} & 8.2\,\std{0.0} & 15.8\,\std{0.0} & 20.9\,\std{0.0} & 8.9\,\std{0.0} & 2.1\,\std{0.0} & 10.6\,\std{0.0} \\
(a3)~$+$~Qwen3-Reranker-4B & 29.7\,\std{0.0} & 13.1\,\std{0.0} & 9.1\,\std{0.0} & 17.3\,\std{0.0} & 22.8\,\std{0.0} & 9.8\,\std{0.0} & 2.4\,\std{0.0} & 11.7\,\std{0.0} \\
(a4)~$+$~\method{}-4B ($\mathrm{M0}$) & \textbf{32.4}\,\std{0.0} & \textbf{14.5}\,\std{0.0} & \textbf{10.5}\,\std{0.1} & \textbf{19.1}\,\std{0.0} & \textbf{23.8}\,\std{0.0} & \textbf{10.2}\,\std{0.0} & \textbf{2.7}\,\std{0.0} & \textbf{12.2}\,\std{0.0} \\
\midrule
(b1)~Qwen3-Embedding-4B & 19.6\,\std{0.0} & 1.8\,\std{0.0} & 7.9\,\std{0.0} & 9.8\,\std{0.0} & 18.1\,\std{0.0} & 1.2\,\std{0.0} & 2.0\,\std{0.0} & 7.1\,\std{0.0} \\
(b2)~$+$~Jina-Reranker-v3 & 28.9\,\std{0.0} & 6.0\,\std{0.0} & 7.9\,\std{0.0} & 14.3\,\std{0.0} & 24.9\,\std{0.0} & 4.7\,\std{0.0} & 2.2\,\std{0.0} & 10.6\,\std{0.0} \\
(b3)~$+$~Qwen3-Reranker-4B & 33.0\,\std{0.1} & 6.5\,\std{0.0} & 8.1\,\std{0.0} & 15.9\,\std{0.0} & 27.1\,\std{0.1} & 4.6\,\std{0.1} & 2.1\,\std{0.0} & 11.2\,\std{0.0} \\
(b4)~$+$~\method{}-4B ($\mathrm{M0}$) & \textbf{37.5}\,\std{0.1} & \textbf{7.6}\,\std{0.0} & \textbf{9.8}\,\std{0.1} & \textbf{18.3}\,\std{0.0} & \textbf{30.2}\,\std{0.1} & \textbf{5.3}\,\std{0.0} & \textbf{2.5}\,\std{0.0} & \textbf{12.7}\,\std{0.0} \\
\midrule
(c1)~AgentIR-4B & 32.0\,\std{0.0} & 11.1\,\std{0.0} & 9.8\,\std{0.0} & 17.6\,\std{0.0} & 24.2\,\std{0.0} & 8.6\,\std{0.0} & 2.6\,\std{0.0} & 11.8\,\std{0.0} \\
(c2)~$+$~Jina-Reranker-v3 & 30.4\,\std{0.0} & 12.2\,\std{0.0} & 9.5\,\std{0.0} & 17.4\,\std{0.0} & 25.6\,\std{0.0} & 9.3\,\std{0.0} & 2.6\,\std{0.0} & 12.5\,\std{0.0} \\
(c3)~$+$~Qwen3-Reranker-4B & 36.4\,\std{0.1} & 14.1\,\std{0.0} & 9.1\,\std{0.0} & 19.9\,\std{0.0} & 29.3\,\std{0.1} & 11.1\,\std{0.0} & 2.6\,\std{0.0} & 14.3\,\std{0.0} \\
(c4)~$+$~\method{}-4B ($\mathrm{M0}$) & \textbf{40.2}\,\std{0.0} & \textbf{15.5}\,\std{0.1} & \textbf{10.6}\,\std{0.1} & \textbf{22.1}\,\std{0.0} & \textbf{31.3}\,\std{0.0} & \textbf{11.6}\,\std{0.0} & \textbf{2.7}\,\std{0.0} & \textbf{15.2}\,\std{0.0} \\
\midrule
(d1)~Nemotron-3-Embed-8B & 27.0\,\std{0.0} & 4.8\,\std{0.0} & \textbf{11.6}\,\std{0.0} & 14.4\,\std{0.0} & 24.1\,\std{0.0} & 3.4\,\std{0.0} & \textbf{3.2}\,\std{0.0} & 10.2\,\std{0.0} \\
(d2)~$+$~Jina-Reranker-v3 & 30.1\,\std{0.0} & 10.3\,\std{0.0} & 7.6\,\std{0.0} & 16.0\,\std{0.0} & 24.9\,\std{0.0} & 8.8\,\std{0.0} & 1.9\,\std{0.0} & 11.9\,\std{0.0} \\
(d3)~$+$~Qwen3-Reranker-4B & 35.0\,\std{0.0} & 11.6\,\std{0.0} & 9.0\,\std{0.1} & 18.5\,\std{0.0} & 28.7\,\std{0.1} & 9.4\,\std{0.0} & 2.5\,\std{0.1} & 13.5\,\std{0.0} \\
(d4)~$+$~\method{}-4B ($\mathrm{M0}$) & \textbf{40.2}\,\std{0.1} & \textbf{13.5}\,\std{0.0} & 11.1\,\std{0.1} & \textbf{21.6}\,\std{0.1} & \textbf{32.4}\,\std{0.2} & \textbf{10.4}\,\std{0.1} & 2.9\,\std{0.0} & \textbf{15.2}\,\std{0.1} \\
\bottomrule
\end{tabular}%
}
\end{table}

\FloatBarrier

\section{Per-benchmark agentic results}\label{app:agentic-benchmarks}

\Cref{tab:agentic-bcp,tab:agentic-bright,tab:agentic-trqa} report results underlying \Cref{tab:agentic-main}, with BM25 on BCP, \mbox{Nemotron-3-Embed-8B} on BRIGHT, and \mbox{AgentIR-4B} on TRQA.
Each agent is evaluated without reranking and with each 4B reranker.
Aggregation and testing follow \Cref{app:stats}.

The benchmark breakdown shows effects obscured by overall averages.
In particular, Qwen3-Reranker-4B lowers mean TRQA nDCG@10 for four agents, while \method{} improves all five.
Document selection and caching also affect API cost (\Cref{app:cost}).

\begin{table}[!tp]
\centering
\setlength{\tabcolsep}{5pt}
\small
\caption{\textbf{Agentic retrieval on BCP}, with BM25 as the first-stage retriever.}
\label{tab:agentic-bcp}
\fitappendixtable{%
\begin{tabular}{@{}l lllll@{}}
\toprule
\textbf{Agent LLM} & \textbf{nDCG@10}$\uparrow$ & \textbf{Recall@10}$\uparrow$ & \multicolumn{3}{c}{\textbf{Avg. per query stats}} \\
\cmidrule(lr){4-6}
$+$\textbf{Reranker} & & & \textbf{Searches}$\downarrow$ & \textbf{LLM turns}$\downarrow$ & \textbf{Cost (\textcent)}$\downarrow$ \\
\midrule
(a1)~Nemotron-3 Ultra & 66.9~\std{1.3} & 68.6~\std{1.0} & 14.9~\std{0.4} & 16.0~\std{0.4} & 35.0~\std{1.8} \\
(a2)~$+$~Qwen3-Reranker-4B & 69.7~\std{1.1} & 72.2~\std{1.5}$^{\dagger}$ & 13.1~\std{0.3}$^{\dagger}$ & 14.1~\std{0.3}$^{\dagger}$ & 30.8~\std{0.8}$^{\dagger}$ \\
(a3)~$+$~\method{}-4B ($\mathrm{M3}$) & \textbf{72.6}~\std{1.2}$^{\dagger*}$ & \textbf{75.2}~\std{0.9}$^{\dagger}$ & \textbf{11.6}~\std{0.3}$^{\dagger*}$ & \textbf{12.6}~\std{0.3}$^{\dagger*}$ & \textbf{25.9}~\std{1.4}$^{\dagger*}$ \\
\midrule
(b1)~MiniMax M3 & 76.7~\std{0.7} & 79.8~\std{0.4} & 17.0~\std{0.4} & 9.1~\std{0.2} & 9.5~\std{1.2} \\
(b2)~$+$~Qwen3-Reranker-4B & 80.4~\std{1.4}$^{\dagger}$ & 84.0~\std{1.3}$^{\dagger}$ & 14.7~\std{0.2}$^{\dagger}$ & 8.1~\std{0.1}$^{\dagger}$ & 8.7~\std{0.7} \\
(b3)~$+$~\method{}-4B ($\mathrm{M3}$) & \textbf{82.5}~\std{1.6}$^{\dagger}$ & \textbf{86.0}~\std{1.4}$^{\dagger}$ & \textbf{14.1}~\std{0.2}$^{\dagger*}$ & \textbf{7.9}~\std{0.1}$^{\dagger*}$ & \textbf{8.4}~\std{0.5} \\
\midrule
(c1)~Gemini 3.6 Flash & 72.3~\std{0.6} & 73.7~\std{0.6} & 8.1~\std{0.1} & 9.1~\std{0.1} & 13.7~\std{0.4} \\
(c2)~$+$~Qwen3-Reranker-4B & 78.3~\std{0.9}$^{\dagger}$ & 80.4~\std{0.9}$^{\dagger}$ & 7.0~\std{0.2}$^{\dagger}$ & 8.0~\std{0.2}$^{\dagger}$ & 12.2~\std{0.4}$^{\dagger}$ \\
(c3)~$+$~\method{}-4B ($\mathrm{M2}$) & \textbf{82.1}~\std{0.5}$^{\dagger*}$ & \textbf{84.6}~\std{0.5}$^{\dagger*}$ & \textbf{6.7}~\std{0.1}$^{\dagger*}$ & \textbf{7.7}~\std{0.1}$^{\dagger*}$ & \textbf{11.7}~\std{0.6}$^{\dagger}$ \\
\midrule
(d1)~GPT-5.6 Terra & 83.0~\std{1.0} & 85.6~\std{0.9} & 20.8~\std{0.4} & 4.5~\std{0.0} & 47.2~\std{1.7} \\
(d2)~$+$~Qwen3-Reranker-4B & 85.2~\std{0.8} & 87.4~\std{0.8} & 18.4~\std{0.2}$^{\dagger}$ & 4.2~\std{0.1}$^{\dagger}$ & 39.5~\std{0.7}$^{\dagger}$ \\
(d3)~$+$~\method{}-4B ($\mathrm{M2}$) & \textbf{86.2}~\std{0.8}$^{\dagger}$ & \textbf{88.7}~\std{0.8}$^{\dagger}$ & \textbf{17.5}~\std{0.5}$^{\dagger}$ & \textbf{4.1}~\std{0.1}$^{\dagger}$ & \textbf{38.3}~\std{2.0}$^{\dagger}$ \\
\midrule
(e1)~Claude Sonnet 5 & 78.7~\std{0.6} & 82.1~\std{0.5} & 15.9~\std{0.6} & 8.9~\std{0.3} & 63.8~\std{3.3} \\
(e2)~$+$~Qwen3-Reranker-4B & 80.2~\std{0.9} & 83.1~\std{0.9} & 12.7~\std{0.2}$^{\dagger}$ & 7.4~\std{0.1}$^{\dagger}$ & 48.9~\std{0.8}$^{\dagger}$ \\
(e3)~$+$~\method{}-4B ($\mathrm{M2}$) & \textbf{82.2}~\std{0.6}$^{\dagger}$ & \textbf{85.2}~\std{0.6}$^{\dagger*}$ & \textbf{11.8}~\std{0.4}$^{\dagger*}$ & \textbf{7.0}~\std{0.2}$^{\dagger}$ & \textbf{45.6}~\std{2.2}$^{\dagger}$ \\
\bottomrule
\end{tabular}%
}
\end{table}

\begin{table}[!tp]
\centering
\setlength{\tabcolsep}{5pt}
\small
\caption{\textbf{Agentic retrieval on BRIGHT}, with \mbox{Nemotron-3-Embed-8B} as the first-stage retriever.}
\label{tab:agentic-bright}
\fitappendixtable{%
\begin{tabular}{@{}l lllll@{}}
\toprule
\textbf{Agent LLM} & \textbf{nDCG@10}$\uparrow$ & \textbf{Recall@10}$\uparrow$ & \multicolumn{3}{c}{\textbf{Avg. per query stats}} \\
\cmidrule(lr){4-6}
$+$\textbf{Reranker} & & & \textbf{Searches}$\downarrow$ & \textbf{LLM turns}$\downarrow$ & \textbf{Cost (\textcent)}$\downarrow$ \\
\midrule
(a1)~Nemotron-3 Ultra & 41.1~\std{0.4} & 45.4~\std{0.4} & 8.1~\std{0.1} & 9.1~\std{0.1} & 9.2~\std{0.5} \\
(a2)~$+$~Qwen3-Reranker-4B & 42.5~\std{0.6}$^{\dagger}$ & \textbf{47.1}~\std{0.6}$^{\dagger}$ & 7.5~\std{0.0}$^{\dagger}$ & 8.5~\std{0.0}$^{\dagger}$ & \textbf{8.7}~\std{0.5} \\
(a3)~$+$~\method{}-4B ($\mathrm{M3}$) & \textbf{42.7}~\std{0.3}$^{\dagger}$ & \textbf{47.1}~\std{0.6}$^{\dagger}$ & \textbf{7.1}~\std{0.0}$^{\dagger*}$ & \textbf{8.1}~\std{0.0}$^{\dagger*}$ & \textbf{8.7}~\std{0.5} \\
\midrule
(b1)~MiniMax M3 & 41.7~\std{0.3} & 46.4~\std{0.5} & 11.9~\std{0.1} & 6.6~\std{0.1} & 3.7~\std{0.2} \\
(b2)~$+$~Qwen3-Reranker-4B & 43.0~\std{0.4}$^{\dagger}$ & 48.1~\std{0.3}$^{\dagger}$ & 10.8~\std{0.2}$^{\dagger}$ & 6.1~\std{0.1}$^{\dagger}$ & \textbf{3.5}~\std{0.2} \\
(b3)~$+$~\method{}-4B ($\mathrm{M3}$) & \textbf{43.2}~\std{0.5}$^{\dagger}$ & \textbf{48.2}~\std{0.4}$^{\dagger}$ & \textbf{10.4}~\std{0.1}$^{\dagger*}$ & \textbf{5.9}~\std{0.0}$^{\dagger}$ & 3.7~\std{0.2} \\
\midrule
(c1)~Gemini 3.6 Flash & 43.3~\std{0.2} & 48.1~\std{0.5} & 7.0~\std{0.1} & 8.0~\std{0.0} & 8.3~\std{0.4} \\
(c2)~$+$~Qwen3-Reranker-4B & 45.3~\std{0.2}$^{\dagger}$ & 50.3~\std{0.2}$^{\dagger}$ & 6.5~\std{0.0}$^{\dagger}$ & 7.5~\std{0.0}$^{\dagger}$ & \textbf{8.1}~\std{0.3} \\
(c3)~$+$~\method{}-4B ($\mathrm{M2}$) & \textbf{45.4}~\std{0.3}$^{\dagger}$ & \textbf{51.0}~\std{0.4}$^{\dagger}$ & \textbf{6.2}~\std{0.0}$^{\dagger*}$ & \textbf{7.2}~\std{0.0}$^{\dagger*}$ & 8.3~\std{0.3} \\
\midrule
(d1)~GPT-5.6 Terra & 44.0~\std{0.3} & 48.9~\std{0.3} & 9.1~\std{0.1} & 3.0~\std{0.0} & \textbf{11.8}~\std{0.1} \\
(d2)~$+$~Qwen3-Reranker-4B & 45.0~\std{0.3}$^{\dagger}$ & 49.7~\std{0.5} & 8.8~\std{0.0}$^{\dagger}$ & \textbf{2.9}~\std{0.0}$^{\dagger}$ & \textbf{11.8}~\std{0.0} \\
(d3)~$+$~\method{}-4B ($\mathrm{M2}$) & \textbf{45.1}~\std{0.2}$^{\dagger}$ & \textbf{50.4}~\std{0.2}$^{\dagger}$ & \textbf{8.4}~\std{0.0}$^{\dagger*}$ & \textbf{2.9}~\std{0.0}$^{\dagger*}$ & 12.4~\std{0.1}$^{\dagger*}$ \\
\midrule
(e1)~Claude Sonnet 5 & 42.1~\std{0.1} & 47.2~\std{0.4} & 8.3~\std{0.0} & 5.2~\std{0.0} & \textbf{16.9}~\std{0.3} \\
(e2)~$+$~Qwen3-Reranker-4B & 42.5~\std{0.3} & 47.3~\std{0.5} & 7.8~\std{0.1}$^{\dagger}$ & 4.9~\std{0.0}$^{\dagger}$ & 17.2~\std{0.2} \\
(e3)~$+$~\method{}-4B ($\mathrm{M2}$) & \textbf{43.2}~\std{0.4}$^{\dagger}$ & \textbf{48.4}~\std{0.5}$^{\dagger}$ & \textbf{7.1}~\std{0.0}$^{\dagger*}$ & \textbf{4.7}~\std{0.0}$^{\dagger*}$ & 17.1~\std{0.2} \\
\bottomrule
\end{tabular}%
}
\end{table}

\begin{table}[!tp]
\centering
\setlength{\tabcolsep}{5pt}
\small
\caption{\textbf{Agentic retrieval on TRQA}, with \mbox{AgentIR-4B} as the first-stage retriever.}
\label{tab:agentic-trqa}
\fitappendixtable{%
\begin{tabular}{@{}l lllll@{}}
\toprule
\textbf{Agent LLM} & \textbf{nDCG@10}$\uparrow$ & \textbf{Recall@10}$\uparrow$ & \multicolumn{3}{c}{\textbf{Avg. per query stats}} \\
\cmidrule(lr){4-6}
$+$\textbf{Reranker} & & & \textbf{Searches}$\downarrow$ & \textbf{LLM turns}$\downarrow$ & \textbf{Cost (\textcent)}$\downarrow$ \\
\midrule
(a1)~Nemotron-3 Ultra & 36.7~\std{0.3} & 24.9~\std{0.2} & 10.4~\std{0.1} & 11.4~\std{0.1} & 12.2~\std{0.9} \\
(a2)~$+$~Qwen3-Reranker-4B & 35.8~\std{0.6} & 24.9~\std{0.4} & 10.3~\std{0.2} & 11.3~\std{0.2} & 11.8~\std{0.6} \\
(a3)~$+$~\method{}-4B ($\mathrm{M3}$) & \textbf{38.3}~\std{0.4}$^{\dagger*}$ & \textbf{26.1}~\std{0.4}$^{\dagger*}$ & \textbf{9.1}~\std{0.1}$^{\dagger*}$ & \textbf{10.1}~\std{0.1}$^{\dagger*}$ & \textbf{10.4}~\std{0.5} \\
\midrule
(b1)~MiniMax M3 & 34.8~\std{0.9} & 24.3~\std{0.6} & 21.4~\std{0.3} & 10.6~\std{0.2} & 6.3~\std{0.5} \\
(b2)~$+$~Qwen3-Reranker-4B & 34.1~\std{0.4} & 24.4~\std{0.4} & 19.8~\std{0.4}$^{\dagger}$ & 9.9~\std{0.2}$^{\dagger}$ & 5.5~\std{0.5} \\
(b3)~$+$~\method{}-4B ($\mathrm{M3}$) & \textbf{37.4}~\std{0.6}$^{\dagger*}$ & \textbf{26.1}~\std{0.1}$^{\dagger*}$ & \textbf{18.5}~\std{0.3}$^{\dagger*}$ & \textbf{9.3}~\std{0.2}$^{\dagger*}$ & \textbf{5.4}~\std{0.4} \\
\midrule
(c1)~Gemini 3.6 Flash & 40.6~\std{0.2} & 26.8~\std{0.2} & 15.1~\std{0.5} & 16.0~\std{0.5} & 19.8~\std{1.3} \\
(c2)~$+$~Qwen3-Reranker-4B & 40.1~\std{0.5} & 27.0~\std{0.3} & 14.6~\std{0.3} & 15.6~\std{0.3} & \textbf{18.3}~\std{0.9} \\
(c3)~$+$~\method{}-4B ($\mathrm{M2}$) & \textbf{42.2}~\std{0.5}$^{\dagger*}$ & \textbf{28.0}~\std{0.4}$^{\dagger*}$ & \textbf{14.2}~\std{0.3} & \textbf{15.2}~\std{0.2} & 19.6~\std{0.7} \\
\midrule
(d1)~GPT-5.6 Terra & 43.3~\std{0.3} & 27.9~\std{0.5} & 25.5~\std{0.3} & 5.0~\std{0.0} & 25.7~\std{0.3} \\
(d2)~$+$~Qwen3-Reranker-4B & 42.7~\std{0.6} & 27.7~\std{0.4} & \textbf{23.7}~\std{0.2}$^{\dagger}$ & \textbf{4.8}~\std{0.0}$^{\dagger}$ & \textbf{23.8}~\std{0.1}$^{\dagger}$ \\
(d3)~$+$~\method{}-4B ($\mathrm{M2}$) & \textbf{44.1}~\std{0.4}$^{*}$ & \textbf{28.6}~\std{0.3} & 24.0~\std{0.3}$^{\dagger}$ & \textbf{4.8}~\std{0.0}$^{\dagger}$ & 25.0~\std{0.3}$^{\dagger*}$ \\
\midrule
(e1)~Claude Sonnet 5 & 36.3~\std{0.5} & 25.2~\std{0.5} & 16.7~\std{0.1} & 7.8~\std{0.1} & 28.7~\std{0.4} \\
(e2)~$+$~Qwen3-Reranker-4B & 36.5~\std{0.4} & 25.6~\std{0.2} & 16.3~\std{0.3} & 7.4~\std{0.1}$^{\dagger}$ & \textbf{27.8}~\std{0.3} \\
(e3)~$+$~\method{}-4B ($\mathrm{M2}$) & \textbf{38.6}~\std{0.6}$^{\dagger*}$ & \textbf{26.5}~\std{0.4}$^{\dagger*}$ & \textbf{15.9}~\std{0.2}$^{\dagger}$ & \textbf{7.3}~\std{0.1}$^{\dagger}$ & 28.1~\std{0.5} \\
\bottomrule
\end{tabular}%
}
\end{table}

\FloatBarrier

\section{Cost analysis}\label{app:cost}

We report the agent LLM's API cost per query.
Retrieval and reranker serving costs are excluded from all experiments, including those in the Analysis section (\Cref{sec:ablations}).
Reported savings concern agent API charges, not total system cost; excluded costs may matter more for cheaper agents.

Costs are computed from recorded token usage using the rates in \Cref{tab:pricing}, with reasoning tokens counted as output.
Public list prices are as of September 2026.
Lightning uses the experiment-recorded rates for its OpenRouter route: \$0.08 for input, \$0.04 for cache reads, and \$0.20 for output per million tokens.
Local inference is valued at the same rates, as a token-priced estimate.
\mbox{GPT-5.5} judging uses recorded OpenAI rates of \$5 for input and \$30 for output per million tokens; its published cache-read rate\footnote{\url{https://developers.openai.com/api/docs/models/gpt-5.5}} is \$0.50.
No cache usage was recorded for the judge.
MiniMax M3 has no recorded cache writes, so its cost calculation requires no cache-write rate.
Because the conversation is resubmitted at each turn, a document can incur input charges repeatedly after it is first retrieved.
Fewer searches and shorter trajectories can therefore reduce cost, but the number of calls alone does not determine the savings.

The token-priced totals in \Cref{tab:pricing} sum each recorded run's per-query cost multiplied by its number of questions, plus retained answer-judge usage for both Piika studies.
Runs reused across studies are counted once using their source-result paths.
These totals exclude omitted experiments, unrecorded attempts, training costs, and judging outside those studies.
Piika contributes 36 ClimbMix runs and 30 BCP context-control runs (\Cref{app:piika-climbmix,app:piika-context}), totaling 8{,}052 question executions.
Its recorded usage is valued at \$50.30 for GPT-5.6 Luna, \$123.92 for Nemotron-3.5 Lightning, and \$64.21 for 7{,}802 \mbox{GPT-5.5} judgments through OpenAI.
The remaining 250 executions incur no judge cost.
The Lightning total includes local inference valued at API token rates, so the combined total is a token-priced accounting estimate.
Piika cost comparisons are omitted because serving conditions affected billed cost: Lightning runs switched between API providers and local serving under heavy throttling, and each Luna reranking configuration ran on its own vLLM server under a different load, so queuing delays changed agent-side prompt-cache reuse.
Search counts are reported throughout the agentic evaluations as a measure of retrieval effort.

Two additional factors affect the cost of a trajectory.
First, document length and retrieval timing determine how much text appears in later prompts: a long document retrieved early contributes more repeated input than one retrieved near the end.
Second, caching determines how much of that input receives a discounted rate.
With implicit caching, cache availability can vary with endpoint workload and cache eviction, changing the billed cost even for similar token volumes.
The reranker influences which documents are returned, but does not control the provider's cache.
Our measurements do not isolate the contributions of caching and document timing to cost variance.

Averaged over benchmarks, \method{} uses significantly fewer search calls than Qwen3-Reranker-4B for every agent, but has significantly lower API cost only for Nemotron-3 Ultra, by approximately 12\%.
Sonnet has a 3\% mean saving that does not remain significant after correction.
The other three agents' cost differences are also nonsignificant, including MiniMax M3's 0.069-cent saving ($p=1.000$; \Cref{tab:agentic-significance}).
These five runs do not establish equal costs.

Benchmark-specific outcomes are more mixed.
Relative to Qwen3-Reranker-4B, \method{} has higher mean API cost in seven cases: Nemotron-3 Ultra, MiniMax M3, Gemini 3.6 Flash, and GPT-5.6 Terra on BRIGHT, and Gemini 3.6 Flash, GPT-5.6 Terra, and Claude Sonnet 5 on TRQA.
Only Terra's two increases remain significant after correction (\Cref{tab:agentic-significance}).
MiniMax M3's cost differences are nonsignificant on all three benchmarks ($p=0.969$ on BRIGHT; $p=1.000$ on BCP and TRQA).
Thus the aggregate quality and search-effort gains should not be interpreted as uniform cost improvements.
Search calls provide a direct measure of retrieval effort, while API cost measures the combined effect of trajectory length, document content, and provider pricing.

\FloatBarrier

\section{Latency and reranking depth}\label{app:depth-latency}

\Cref{tab:ablation-depth} examines whether reduced search effort translates into lower end-to-end latency for Nemotron-3 Ultra on BCP with BM25.
Each search retrieves and reranks $d$ candidates, for $d\in\{50,100\}$; the no-reranking baseline retrieves 100.
Output depth remains ten unseen documents by default.
The reranker runs on one A100 GPU, with two questions processed concurrently.
Latency is elapsed time per query, including retrieval, reranking, LLM calls, and any recovery, rather than total run time divided by the number of questions.

\begin{table}[!t]
\centering
\setlength{\tabcolsep}{5pt}
\small
\caption{\textbf{Reranking depth} on BCP with Nemotron-3 Ultra and BM25.
Each depth compares the two rerankers with the shared no-reranker baseline; row definitions are in \Cref{app:depth-latency}.
Tests use this study's correction family (\Cref{app:stats}); marks may differ for shared runs in \Cref{tab:agentic-bcp,tab:agentic-bright,tab:agentic-trqa}.}
\label{tab:ablation-depth}
\vspace{2pt}
\fitappendixtable{%
\begin{tabular}{@{}l lllll@{}}
\toprule
& & \multicolumn{2}{c}{\textbf{Depth 50}} & \multicolumn{2}{c}{\textbf{Depth 100}} \\
\cmidrule(lr){3-4}\cmidrule(lr){5-6}
& \textbf{No reranker} & \begin{tabular}[b]{@{}c@{}}\textbf{Qwen3-}\\\textbf{Reranker-4B}\end{tabular} & \begin{tabular}[b]{@{}c@{}}\textbf{\method{}-4B ($\mathrm{M3}$)}\end{tabular} & \begin{tabular}[b]{@{}c@{}}\textbf{Qwen3-}\\\textbf{Reranker-4B}\end{tabular} & \begin{tabular}[b]{@{}c@{}}\textbf{\method{}-4B ($\mathrm{M3}$)}\end{tabular} \\
\midrule
\multicolumn{6}{@{}l}{\textit{Quality}} \\
\quad nDCG@10$\uparrow$ & 66.9~\std{1.3} & 69.0~\std{1.0} & \textbf{70.5}~\std{1.1}$^{\dagger}$ & 69.7~\std{1.1}$^{\dagger}$ & \textbf{72.6}~\std{1.2}$^{\dagger*}$ \\
\quad Recall@10$\uparrow$ & 68.6~\std{1.0} & 71.3~\std{1.3}$^{\dagger}$ & \textbf{72.5}~\std{1.5}$^{\dagger}$ & 72.2~\std{1.5}$^{\dagger}$ & \textbf{75.2}~\std{0.9}$^{\dagger*}$ \\
\midrule
\multicolumn{6}{@{}l}{\textit{Avg. per query stats}} \\
\quad Searches$\downarrow$ & 14.9~\std{0.4} & 13.9~\std{0.1}$^{\dagger}$ & \textbf{13.3}~\std{0.4}$^{\dagger}$ & 13.1~\std{0.3}$^{\dagger}$ & \textbf{11.6}~\std{0.3}$^{\dagger*}$ \\
\quad LLM turns$\downarrow$ & 16.0~\std{0.4} & 14.9~\std{0.1}$^{\dagger}$ & \textbf{14.4}~\std{0.4}$^{\dagger}$ & 14.1~\std{0.3}$^{\dagger}$ & \textbf{12.6}~\std{0.3}$^{\dagger*}$ \\
\quad Cost (\textcent)$\downarrow$ & 35.0~\std{1.8} & 33.1~\std{0.7} & \textbf{31.9}~\std{2.0} & 30.8~\std{0.8}$^{\dagger}$ & \textbf{25.9}~\std{1.4}$^{\dagger*}$ \\
\quad End-to-end latency (s)$\downarrow$ & \textbf{71.7}~\std{3.1} & 90.8~\std{1.0}$^{\dagger}$ & 84.1~\std{2.8}$^{\dagger*}$ & 99.5~\std{4.2}$^{\dagger}$ & 91.6~\std{4.5}$^{\dagger}$ \\
\bottomrule
\end{tabular}%
}
\end{table}

Both rerankers increase latency relative to no reranking.
Compared with Qwen3-Reranker-4B, \method{} reduces mean end-to-end latency from 90.8 to 84.1\,s at depth 50 ($p=0.025$) and from 99.5 to 91.6\,s at depth 100 ($p=0.110$).
Corresponding nDCG@10 values increase from 69.0 to 70.5 ($p=0.299$) and from 69.7 to 72.6 ($p=0.025$).
At depth 100, Recall@10 also improves from 72.2 to 75.2 ($p=0.038$).
Thus the depth-50 latency reduction and depth-100 gains in both quality metrics are significant under the 36-comparison correction defined in \Cref{app:stats}.
Relative to no reranking, depth 100 adds approximately 0.29 nDCG@10 pp per second of added latency for \method{}, compared with 0.10 for Qwen3-Reranker-4B.

\method{} reduces mean LLM turns by 0.6 at depth 50 ($p=0.140$) and 1.5 at depth 100 ($p=0.001$).
These observations are consistent with shorter trajectories reducing end-to-end work, although they do not isolate the contribution of each component to latency.

\FloatBarrier

\section{Small vs. large agents}\label{app:size-cost}

\Cref{tab:ablation-size} reports five-run results on BCP with BM25 and reranking depth 100 for three pairs: GPT-5.6 Luna/Terra, Gemini 3.5 Flash-Lite/3.6 Flash, and Claude Haiku 4.5/Sonnet 5.
\method{} uses the original-question mode $\mathrm{M2}$ for all six agents.
Relative cost is API cost divided by the mean API cost of the larger model without reranking in the same family.
This fixed normalization also scales the reported standard deviations; the baseline mean is 100\%, but its standard deviation can be nonzero.

\method{} improves mean nDCG@10 over no reranking by 3.3--12.6 pp, and over Qwen3-Reranker-4B by 1.0--6.3 pp. The gains over no reranking are significant for all six agents under the reported tests.
After correction across the 51 size-study comparisons, nDCG@10 gains over Qwen3-Reranker-4B are significant for Flash-Lite ($p=0.019$), Flash ($p=0.002$), Haiku ($p=0.014$), and Sonnet ($p=0.036$).
Recall@10 gains remain significant for Flash-Lite, Flash, Haiku, and Sonnet.
Within each family, the smaller agent gains at least as much as the larger one relative to no reranking.
These are comparisons between deployed model tiers, not isolated changes in parameter count; the Gemini and Claude pairs also differ in version or generation.

Let $S_0$ and $L_0$ denote the smaller and larger agents' mean nDCG@10 without reranking, and let $S_r$ denote the smaller agent's mean with reranker $r$.
Gap closure relative to the larger agent without reranking is $(S_r-S_0)/(L_0-S_0)$.
Qwen3-Reranker-4B closes 46\%, 46\%, and 26\% of the GPT, Gemini, and Claude gaps, respectively; \method{} closes 73\%, 74\%, and 51\%.
For \method{}, the remaining GPT and Gemini gaps are 1.4 pp ($p=0.331$) and 3.6 pp ($p=0.025$); the nonsignificant GPT difference does not establish equal performance.

\begin{table}[!tp]
\centering
\setlength{\tabcolsep}{4pt}
\small
\caption{\textbf{Small versus large agents} on BCP with BM25.
Relative cost is normalized to the larger agent without reranking in each family.
Tests use this study's correction family (\Cref{app:stats}); marks may differ for shared runs in \Cref{tab:agentic-bcp,tab:agentic-bright,tab:agentic-trqa}.}
\label{tab:ablation-size}
\fitappendixtable{%
\begin{tabular}{@{}l lllll@{}}
\toprule
\textbf{Agent LLM} & \textbf{nDCG@10}$\uparrow$ & \textbf{Recall@10}$\uparrow$ & \multicolumn{2}{c}{\textbf{Avg. per query stats}} & \textbf{Rel. cost (\%)} \\
\cmidrule(lr){4-5}
$+$\textbf{Reranker} & & & \textbf{Searches}$\downarrow$ & \textbf{Cost (\textcent)}$\downarrow$ & \\
\midrule
(a1)~GPT-5.6 Luna (small) & 77.9~\std{0.7} & 81.3~\std{0.7} & 19.0~\std{0.5} & 3.8~\std{0.2} & 8.0~\std{0.4} \\
(a2)~$+$~Qwen3-Reranker-4B & 80.3~\std{0.4} & 82.7~\std{0.4} & 17.0~\std{0.5} & 3.4~\std{0.1} & 7.1~\std{0.2} \\
(a3)~$+$~\method{}-4B ($\mathrm{M2}$) & \textbf{81.6}~\std{1.0}$^{\dagger}$ & \textbf{84.5}~\std{1.3}$^{\dagger}$ & \textbf{15.4}~\std{0.5}$^{\dagger*}$ & \textbf{3.1}~\std{0.1}$^{\dagger}$ & \textbf{6.6}~\std{0.3}$^{\dagger}$ \\
\addlinespace[3pt]
(b1)~GPT-5.6 Terra (large) & 83.0~\std{1.0} & 85.6~\std{0.9} & 20.8~\std{0.4} & 47.2~\std{1.7} & 100.0~\std{3.7} \\
(b2)~$+$~Qwen3-Reranker-4B & 85.2~\std{0.8} & 87.4~\std{0.8} & 18.4~\std{0.2} & 39.5~\std{0.7} & 83.7~\std{1.6} \\
(b3)~$+$~\method{}-4B ($\mathrm{M2}$) & \textbf{86.2}~\std{0.8}$^{\dagger}$ & \textbf{88.7}~\std{0.8}$^{\dagger}$ & \textbf{17.5}~\std{0.5}$^{\dagger}$ & \textbf{38.3}~\std{2.0}$^{\dagger}$ & \textbf{81.2}~\std{4.3}$^{\dagger}$ \\
\midrule
(c1)~Gemini 3.5 Flash-Lite (small) & 58.5~\std{1.4} & 59.9~\std{1.4} & 6.4~\std{0.1} & 3.8~\std{0.1} & 27.5~\std{1.0} \\
(c2)~$+$~Qwen3-Reranker-4B & 64.9~\std{1.3} & 66.6~\std{0.7} & \textbf{5.2}~\std{0.1} & \textbf{3.1}~\std{0.1} & \textbf{22.7}~\std{0.4} \\
(c3)~$+$~\method{}-4B ($\mathrm{M2}$) & \textbf{68.7}~\std{1.5}$^{\dagger*}$ & \textbf{71.0}~\std{1.4}$^{\dagger*}$ & 5.3~\std{0.1}$^{\dagger}$ & 3.2~\std{0.1}$^{\dagger}$ & 23.6~\std{0.5}$^{\dagger}$ \\
\addlinespace[3pt]
(d1)~Gemini 3.6 Flash (large) & 72.3~\std{0.6} & 73.7~\std{0.6} & 8.1~\std{0.1} & 13.7~\std{0.4} & 100.0~\std{3.2} \\
(d2)~$+$~Qwen3-Reranker-4B & 78.3~\std{0.9} & 80.4~\std{0.9} & 7.0~\std{0.2} & 12.2~\std{0.4} & 89.1~\std{3.3} \\
(d3)~$+$~\method{}-4B ($\mathrm{M2}$) & \textbf{82.1}~\std{0.5}$^{\dagger*}$ & \textbf{84.6}~\std{0.5}$^{\dagger*}$ & \textbf{6.7}~\std{0.1}$^{\dagger*}$ & \textbf{11.7}~\std{0.6}$^{\dagger}$ & \textbf{85.7}~\std{4.6}$^{\dagger}$ \\
\midrule
(e1)~Claude Haiku 4.5 (small) & 54.0~\std{2.0} & 57.1~\std{2.1} & 15.3~\std{0.6} & 18.0~\std{0.8} & 28.2~\std{1.2} \\
(e2)~$+$~Qwen3-Reranker-4B & 60.4~\std{2.1} & 63.5~\std{2.1} & 14.3~\std{0.6} & 17.2~\std{0.7} & 27.0~\std{1.1} \\
(e3)~$+$~\method{}-4B ($\mathrm{M2}$) & \textbf{66.6}~\std{2.2}$^{\dagger*}$ & \textbf{70.5}~\std{2.2}$^{\dagger*}$ & \textbf{14.0}~\std{0.4}$^{\dagger}$ & \textbf{17.1}~\std{0.6} & \textbf{26.8}~\std{1.0} \\
\addlinespace[3pt]
(f1)~Claude Sonnet 5 (large) & 78.7~\std{0.6} & 82.1~\std{0.5} & 15.9~\std{0.6} & 63.8~\std{3.3} & 100.0~\std{5.1} \\
(f2)~$+$~Qwen3-Reranker-4B & 80.2~\std{0.9} & 83.1~\std{0.9} & 12.7~\std{0.2} & 48.9~\std{0.8} & 76.6~\std{1.3} \\
(f3)~$+$~\method{}-4B ($\mathrm{M2}$) & \textbf{82.2}~\std{0.6}$^{\dagger*}$ & \textbf{85.2}~\std{0.6}$^{\dagger*}$ & \textbf{11.8}~\std{0.4}$^{\dagger*}$ & \textbf{45.6}~\std{2.2}$^{\dagger}$ & \textbf{71.4}~\std{3.4}$^{\dagger}$ \\
\bottomrule
\end{tabular}%
}
\end{table}

\FloatBarrier

Using the larger agent with \method{} as the reference replaces $L_0$ with $L_{\mathrm{ACARe}}$, its mean nDCG@10 with \method{}.
Against this fixed target, adding \method{} to the smaller agent closes 44\%, 43\%, and 45\% of its initial gap.
With \method{} on both sides, the smaller agents' nDCG@10 scores are 4.6, 13.4, and 15.5 pp lower, at 8.1\%, 27.6\%, and 37.5\% of their larger counterparts' API costs.
Compared with neither agent using reranking, the gaps shrink by 8\%, 3\%, and 37\%, respectively.
These comparisons are descriptive.

Every agent has lower mean API cost with reranking than without it.
With \method{}, the larger agents' relative costs are 81.2\%, 85.7\%, and 71.4\% for Terra, Flash, and Sonnet; the smaller agents' costs are 6.6\%, 23.6\%, and 26.8\%.
The reduction from Haiku's no-reranking cost is not significant ($p=0.388$).
No API cost difference between the two rerankers remains significant after correction.

The larger agent with \method{} remains the most effective configuration in each family, with nDCG@10 of 86.2, 82.1, and 82.2.
The small-agent results therefore describe a lower-cost alternative with some loss of quality, under the cost accounting in \Cref{app:cost}.

\FloatBarrier

\section{Retriever headroom on BRIGHT}\label{app:headroom}
\suppressfloats[t]

The main BRIGHT experiments show small nDCG@10 differences between \method{} and Qwen3-Reranker-4B (\Cref{tab:agentic-significance}).
\mbox{Nemotron-3-Embed-8B}'s training includes synthetic query--document pairs generated using BRIGHT as a seed corpus.\footnote{\url{https://huggingface.co/nvidia/Nemotron-3-Embed-8B-BF16}}
This corpus-specific supervision may reduce the room for further gains from agentic search and reranking.
The BRIGHT results with this retriever therefore involve benchmark-seeded training data; the comparison below tests a different retriever and training recipe.
We examine this possibility by replacing \mbox{Nemotron-3-Embed-8B} with the general-purpose \mbox{Qwen3-Embedding-4B} \citep{zhang2025qwen3embed} for the Nemotron-3 Ultra agent.
The agent, question sample, five run indices, search interface, and reranking depth remain fixed.
Both use BRIGHT; the Nemotron rows repeat the main results.

\begin{inplacetable}
\centering
\setlength{\tabcolsep}{4pt}
\small
\caption{\textbf{Retriever headroom on BRIGHT} with Nemotron-3 Ultra: \mbox{Nemotron-3-Embed-8B}, used in the main experiments, versus \mbox{Qwen3-Embedding-4B}.}
\label{tab:ablation-headroom}
\fitappendixtable{%
\begin{tabular}{@{}l lllll@{}}
\toprule
\textbf{Retriever} & \textbf{nDCG@10}$\uparrow$ & \textbf{Recall@10}$\uparrow$ & \multicolumn{3}{c}{\textbf{Avg. per query stats}} \\
\cmidrule(lr){4-6}
$+$\textbf{Reranker} & & & \textbf{Searches}$\downarrow$ & \textbf{LLM turns}$\downarrow$ & \textbf{Cost (\textcent)}$\downarrow$ \\
\midrule
(a1)~Nemotron-3-Embed-8B & 41.1~\std{0.4} & 45.4~\std{0.4} & 8.1~\std{0.1} & 9.1~\std{0.1} & 9.2~\std{0.5} \\
(a2)~$+$~Qwen3-Reranker-4B & 42.5~\std{0.6} & \textbf{47.1}~\std{0.6} & 7.5~\std{0.0} & 8.5~\std{0.0} & \textbf{8.7}~\std{0.5} \\
(a3)~$+$~\method{}-4B ($\mathrm{M3}$) & \textbf{42.7}~\std{0.3} & \textbf{47.1}~\std{0.6} & \textbf{7.1}~\std{0.0}$^{*}$ & \textbf{8.1}~\std{0.0}$^{*}$ & \textbf{8.7}~\std{0.5} \\
\midrule
(b1)~Qwen3-Embedding-4B & 40.1~\std{0.4} & 44.2~\std{0.6} & 8.3~\std{0.1} & 9.3~\std{0.1} & 10.7~\std{0.1} \\
(b2)~$+$~Qwen3-Reranker-4B & 41.6~\std{0.3} & 45.9~\std{0.5} & 7.6~\std{0.1} & 8.6~\std{0.1} & \textbf{10.0}~\std{0.4} \\
(b3)~$+$~\method{}-4B ($\mathrm{M3}$) & \textbf{42.5}~\std{0.3}$^{*}$ & \textbf{47.0}~\std{0.4}$^{*}$ & \textbf{7.2}~\std{0.1}$^{*}$ & \textbf{8.2}~\std{0.1}$^{*}$ & 10.3~\std{0.4} \\
\bottomrule
\end{tabular}%
}
\end{inplacetable}

Without reranking, agentic search improves nDCG@10 over one-shot retrieval by 15.8 pp with \mbox{Nemotron-3-Embed-8B} and 18.4 pp with \mbox{Qwen3-Embedding-4B} (\Cref{tab:oneshot-main,tab:ablation-headroom}).
Thus, agentic search provides substantial gains with both retrievers, with a larger gain for \mbox{Qwen3-Embedding-4B}.

With \mbox{Nemotron-3-Embed-8B}, \method{} exceeds Qwen3-Reranker-4B by 0.2 nDCG@10 pp ($p=1.000$), and both have rounded Recall@10 of 47.1 ($p=1.000$).
With \mbox{Qwen3-Embedding-4B}, the differences increase to 0.9 nDCG@10 pp ($p=0.022$) and 1.1 Recall@10 pp ($p=0.033$).
The gain over no reranking also increases, from 1.6 to 2.4 nDCG@10 pp.

Switching to \mbox{Qwen3-Embedding-4B} lowers \method{}'s mean nDCG@10 from 42.7 to 42.5 ($p=1.000$), compared with reductions of 0.9 pp for Qwen3-Reranker-4B ($p=0.122$) and 1.0 without reranking ($p=0.033$).
For the base reranker, Recall@10 also falls by 1.2 pp ($p=0.044$).
These observed differences suggest that \method{} is less sensitive to the retriever change in this setting.
The larger gains with \mbox{Qwen3-Embedding-4B} are consistent with the headroom explanation, but this comparison does not isolate the effect of training data or establish a significant retriever--reranker interaction.

\method{} uses approximately 0.4 fewer searches per query than Qwen3-Reranker-4B with either retriever ($p<0.001$).
Cost differences between the rerankers are not significant in either configuration (both adjusted $p=1.000$).

\FloatBarrier

\section{Aggregation and statistical tests}\label{app:stats}

In the retrieval evaluations, BCP quality metrics average the gold and evidence label sets; BRIGHT and TRQA quality metrics are macro-averages over twelve and three subsets, respectively.
Search calls, LLM turns, and API costs are pooled over the questions within each benchmark.
Overall results average the three benchmark-level values with equal weight.
Tables report means and sample standard deviations over five runs (six for Piika); figures use the same means and standard deviations.

We compare configurations with two-sided Welch $t$-tests on their run-level values and apply Benjamini--Yekutieli (BY) correction~\citep{benjamini2001control} within each comparison family.
BY controls the false discovery rate, the expected fraction of false rejections among rejected hypotheses, under arbitrary dependence between tests, provided the individual $p$-values are valid.
All reported $p$-values are adjusted, with significance defined as $p<0.05$.
Welch tests allow unequal variances but assume independent runs and approximately normal run-level means.
With only five or six runs, these assumptions are difficult to assess.
The tests assess whether differences are consistent across repeated executions of the same questions.
They do not assess generalization to other questions.
A nonsignificant result means that the repeated runs provide insufficient evidence of a difference, not that the systems perform equally.

\paragraph{Main comparisons.}
The aggregate family contains 75 tests: all three pairwise comparisons among no reranking, Qwen3-Reranker-4B, and \method{}, for five agents and five metrics (nDCG@10, Recall@10, searches, LLM turns, and API cost).
The benchmark family contains the corresponding 225 tests over three benchmarks.
Thus claims based on either quality metric account for testing both metrics.
False discovery rate control applies within each specified family, not across the entire paper.
A comparison reused in multiple studies can therefore receive different adjusted $p$-values and significance marks, even when its run-level values are identical.

\Cref{tab:agentic-significance} reports \method{} minus Qwen3-Reranker-4B and the adjusted $p$-values.
Quality differences are pp on the 0--100 scale; cost differences are cents per query.
Negative differences indicate fewer calls or lower cost.
The same $*$ symbol marks significant differences in this table; negative cost differences favor \method{}.
Bold is reserved for best means in result tables.

All five agents retain significant aggregate gains in both quality metrics, fewer searches, and fewer LLM turns.
Aggregate cost is significantly lower only for Nemotron-3 Ultra.
In the benchmark breakdown, nDCG@10 gains remain significant for Ultra and Flash on BCP and for all five agents on TRQA.
Recall@10 gains remain significant for Flash and Sonnet on BCP and for Ultra, M3, Flash, and Sonnet on TRQA.

\paragraph{Analysis comparisons.}
The latency and reranking-depth study (\Cref{app:depth-latency}), small-agent study (\Cref{app:size-cost}), and retriever-headroom study (\Cref{app:headroom}) each have a separate correction family.
The latency and depth study includes 36 tests: all three pairwise comparisons among the no-reranking baseline and two rerankers at each of two depths on six outcomes in \Cref{tab:ablation-depth}: nDCG@10, Recall@10, searches, LLM turns, API cost, and end-to-end latency.
The size study includes 51 tests: \method{} against each baseline for six agents on nDCG@10, Recall@10, searches, and cost (48 tests), plus the three nDCG@10 comparisons between a smaller agent with \method{} and its larger counterpart without reranking.
The retriever-headroom study includes 16 tests: the two rerankers on five metrics under each retriever (ten tests), plus both quality metrics across retrievers for each of the three reranking configurations (six tests).
The context study (\Cref{sec:context-ablation}) has 45 tests: each of four \method{} modes against both baselines, plus Qwen3-Reranker-4B against no reranking, on five metrics.
Comparisons between \method{} context modes remain descriptive.
One-shot results are descriptive (\Cref{app:oneshot}).

\begin{table}[!tp]
\centering
\small
\caption{\textbf{ACARe-4B versus Qwen3-Reranker-4B.}
Cells show the mean difference followed by the adjusted $p$-value; quality differences are pp, and cost differences are cents per query.}
\label{tab:agentic-significance}
\vspace{2pt}
\fitappendixtable{%
\begin{tabular}{@{}l llll@{}}
\toprule
\textbf{Agent LLM, \method{} mode} & \textbf{BCP} & \textbf{BRIGHT} & \textbf{TRQA} & \textbf{Overall} \\
\midrule
\multicolumn{5}{@{}l}{\textit{$\Delta$ nDCG@10 (points)}} \\
\midrule
Nemotron-3 Ultra ($\mathrm{M3}$) & +2.9~{\scriptsize(0.035)}$^{*}$ & +0.2~{\scriptsize(1.000)} & +2.5~{\scriptsize(0.001)}$^{*}$ & +1.9~{\scriptsize(0.005)}$^{*}$ \\
MiniMax M3 ($\mathrm{M3}$) & +2.1~{\scriptsize(0.430)} & +0.2~{\scriptsize(1.000)} & +3.4~{\scriptsize($<$0.001)}$^{*}$ & +1.9~{\scriptsize(0.006)}$^{*}$ \\
Gemini 3.6 Flash ($\mathrm{M2}$) & +3.8~{\scriptsize(0.002)}$^{*}$ & +0.1~{\scriptsize(1.000)} & +2.1~{\scriptsize(0.003)}$^{*}$ & +2.0~{\scriptsize($<$0.001)}$^{*}$ \\
GPT-5.6 Terra ($\mathrm{M2}$) & +1.0~{\scriptsize(0.546)} & +0.1~{\scriptsize(1.000)} & +1.4~{\scriptsize(0.027)}$^{*}$ & +0.9~{\scriptsize(0.022)}$^{*}$ \\
Claude Sonnet 5 ($\mathrm{M2}$) & +2.0~{\scriptsize(0.052)} & +0.7~{\scriptsize(0.155)} & +2.1~{\scriptsize(0.006)}$^{*}$ & +1.6~{\scriptsize($<$0.001)}$^{*}$ \\
\midrule
\multicolumn{5}{@{}l}{\textit{$\Delta$ Recall@10 (points)}} \\
\midrule
Nemotron-3 Ultra ($\mathrm{M3}$) & +3.0~{\scriptsize(0.061)} & \textminus{}0.0~{\scriptsize(1.000)} & +1.3~{\scriptsize(0.012)}$^{*}$ & +1.4~{\scriptsize(0.023)}$^{*}$ \\
MiniMax M3 ($\mathrm{M3}$) & +2.0~{\scriptsize(0.391)} & +0.2~{\scriptsize(1.000)} & +1.7~{\scriptsize(0.006)}$^{*}$ & +1.3~{\scriptsize(0.037)}$^{*}$ \\
Gemini 3.6 Flash ($\mathrm{M2}$) & +4.3~{\scriptsize(0.001)}$^{*}$ & +0.7~{\scriptsize(0.082)} & +1.0~{\scriptsize(0.019)}$^{*}$ & +2.0~{\scriptsize($<$0.001)}$^{*}$ \\
GPT-5.6 Terra ($\mathrm{M2}$) & +1.3~{\scriptsize(0.311)} & +0.6~{\scriptsize(0.367)} & +0.8~{\scriptsize(0.082)} & +0.9~{\scriptsize(0.009)}$^{*}$ \\
Claude Sonnet 5 ($\mathrm{M2}$) & +2.1~{\scriptsize(0.032)}$^{*}$ & +1.0~{\scriptsize(0.077)} & +0.9~{\scriptsize(0.033)}$^{*}$ & +1.3~{\scriptsize($<$0.001)}$^{*}$ \\
\midrule
\multicolumn{5}{@{}l}{\textit{$\Delta$ search calls (per query)}} \\
\midrule
Nemotron-3 Ultra ($\mathrm{M3}$) & \textminus{}1.5~{\scriptsize(0.002)}$^{*}$ & \textminus{}0.4~{\scriptsize($<$0.001)}$^{*}$ & \textminus{}1.2~{\scriptsize($<$0.001)}$^{*}$ & \textminus{}1.0~{\scriptsize($<$0.001)}$^{*}$ \\
MiniMax M3 ($\mathrm{M3}$) & \textminus{}0.7~{\scriptsize(0.006)}$^{*}$ & \textminus{}0.5~{\scriptsize(0.019)}$^{*}$ & \textminus{}1.2~{\scriptsize(0.014)}$^{*}$ & \textminus{}0.8~{\scriptsize($<$0.001)}$^{*}$ \\
Gemini 3.6 Flash ($\mathrm{M2}$) & \textminus{}0.4~{\scriptsize(0.039)}$^{*}$ & \textminus{}0.3~{\scriptsize($<$0.001)}$^{*}$ & \textminus{}0.3~{\scriptsize(0.501)} & \textminus{}0.3~{\scriptsize(0.002)}$^{*}$ \\
GPT-5.6 Terra ($\mathrm{M2}$) & \textminus{}0.9~{\scriptsize(0.077)} & \textminus{}0.4~{\scriptsize($<$0.001)}$^{*}$ & +0.2~{\scriptsize(0.933)} & \textminus{}0.3~{\scriptsize(0.006)}$^{*}$ \\
Claude Sonnet 5 ($\mathrm{M2}$) & \textminus{}0.9~{\scriptsize(0.025)}$^{*}$ & \textminus{}0.6~{\scriptsize($<$0.001)}$^{*}$ & \textminus{}0.5~{\scriptsize(0.200)} & \textminus{}0.7~{\scriptsize(0.002)}$^{*}$ \\
\midrule
\multicolumn{5}{@{}l}{\textit{$\Delta$ LLM turns (per query)}} \\
\midrule
Nemotron-3 Ultra ($\mathrm{M3}$) & \textminus{}1.5~{\scriptsize(0.001)}$^{*}$ & \textminus{}0.4~{\scriptsize($<$0.001)}$^{*}$ & \textminus{}1.2~{\scriptsize($<$0.001)}$^{*}$ & \textminus{}1.0~{\scriptsize($<$0.001)}$^{*}$ \\
MiniMax M3 ($\mathrm{M3}$) & \textminus{}0.2~{\scriptsize(0.019)}$^{*}$ & \textminus{}0.2~{\scriptsize(0.077)} & \textminus{}0.5~{\scriptsize(0.015)}$^{*}$ & \textminus{}0.3~{\scriptsize($<$0.001)}$^{*}$ \\
Gemini 3.6 Flash ($\mathrm{M2}$) & \textminus{}0.4~{\scriptsize(0.044)}$^{*}$ & \textminus{}0.3~{\scriptsize($<$0.001)}$^{*}$ & \textminus{}0.4~{\scriptsize(0.455)} & \textminus{}0.3~{\scriptsize(0.001)}$^{*}$ \\
GPT-5.6 Terra ($\mathrm{M2}$) & \textminus{}0.1~{\scriptsize(0.221)} & \textminus{}0.1~{\scriptsize($<$0.001)}$^{*}$ & +0.0~{\scriptsize(1.000)} & \textminus{}0.1~{\scriptsize(0.011)}$^{*}$ \\
Claude Sonnet 5 ($\mathrm{M2}$) & \textminus{}0.4~{\scriptsize(0.068)} & \textminus{}0.3~{\scriptsize($<$0.001)}$^{*}$ & \textminus{}0.1~{\scriptsize(0.969)} & \textminus{}0.3~{\scriptsize(0.014)}$^{*}$ \\
\midrule
\multicolumn{5}{@{}l}{\textit{$\Delta$ cost (\textcent/query)}} \\
\midrule
Nemotron-3 Ultra ($\mathrm{M3}$) & \textminus{}4.87~{\scriptsize(0.006)}$^{*}$ & +0.02~{\scriptsize(1.000)} & \textminus{}1.35~{\scriptsize(0.068)} & \textminus{}2.07~{\scriptsize(0.004)}$^{*}$ \\
MiniMax M3 ($\mathrm{M3}$) & \textminus{}0.28~{\scriptsize(1.000)} & +0.19~{\scriptsize(0.969)} & \textminus{}0.12~{\scriptsize(1.000)} & \textminus{}0.07~{\scriptsize(1.000)} \\
Gemini 3.6 Flash ($\mathrm{M2}$) & \textminus{}0.46~{\scriptsize(1.000)} & +0.18~{\scriptsize(1.000)} & +1.33~{\scriptsize(0.241)} & +0.35~{\scriptsize(1.000)} \\
GPT-5.6 Terra ($\mathrm{M2}$) & \textminus{}1.20~{\scriptsize(1.000)} & +0.67~{\scriptsize($<$0.001)}$^{*}$ & +1.22~{\scriptsize(0.005)}$^{*}$ & +0.23~{\scriptsize(1.000)} \\
Claude Sonnet 5 ($\mathrm{M2}$) & \textminus{}3.30~{\scriptsize(0.196)} & \textminus{}0.13~{\scriptsize(1.000)} & +0.30~{\scriptsize(1.000)} & \textminus{}1.04~{\scriptsize(0.210)} \\
\bottomrule
\end{tabular}%
}
\end{table}

\paragraph{Piika ClimbMix comparisons.}
We compare \method{} with both baselines on answer accuracy, Recall@all, Precision@all, and searches for both agents, forming one pooled 16-test BY correction family (\Cref{sec:piika}).
Each metric is averaged over all 57 questions before testing across six runs.
Every such comparison is retained regardless of the observed difference.
Comparisons among baselines remain descriptive; monetary cost is excluded from these tests.

\paragraph{Base-reranker context control.}
The separate Piika BCP study (\Cref{app:piika-context}) has its own 16-test BY family: \method{} against no reranking and three Qwen3-Reranker-4B configurations (sub-query only, original question added, and original question plus reasoning added), on answer accuracy, Recall@all, Precision@all, and searches.
Recall and precision each average gold and evidence qrel scores equally.
Metrics average all 200 questions within each of six runs.
All sixteen comparisons are retained.
Comparisons among the base-reranker configurations are descriptive.

\FloatBarrier

\section{Piika: end-to-end answer generation}\label{app:piika}
\suppressfloats[t]
\subsection{Transfer to ClimbMix}\label{app:piika-climbmix}
The ClimbMix study in \Cref{sec:piika} evaluates GPT-5.6 Luna and Nemotron-3.5 Lightning on all 57 BrowseComp-Plus$_{\mathrm{CM}}$ questions over ClimbMix's 553 million documents~\citep{sharifymoghaddam2026cmass}.
Piika exposes \texttt{search} and \texttt{read\_document} tools through a remote Pyserini BM25 REST API~\citep{Lin_etal_SIGIR2021_Pyserini}.
Search returns five excerpts capped at 2{,}000 characters (\texttt{max\_doc\_length}); reranking selects these from the top 100 candidates, while document reads request full text.
Both agents use high reasoning effort and a two-hour query timeout.
Each agent runs six times with each of three retrieval configurations: no reranking, Qwen3-Reranker-4B, and \method{}-4B.

Luna exposes no reasoning trace, so \method{} uses $\mathrm{M2}$.
Lightning exposes its trace, enabling \method{}'s $\mathrm{M3}$.
Lightning has about 30B total parameters and 3B active parameters.\footnote{\href{https://huggingface.co/nvidia/NVIDIA-Nemotron-3.5-Lightning-30B-A3B-BF16}{\nolinkurl{huggingface.co/nvidia/NVIDIA-Nemotron-3.5-Lightning-30B-A3B-BF16}}}
Qwen3-Reranker-4B uses only the sub-query with each candidate.

Before reranking, candidate excerpts are further capped at 300 words for both \method{} and Qwen3-Reranker-4B.
This limit affects reranker inputs, not the excerpts returned to the agent.

The gold-answer judge is GPT-5.5 with low reasoning effort, accessed directly through OpenAI.
All 57 questions contribute to every run's metrics, including 19 Lightning executions that exhausted the 32K-token output budget during reasoning without producing a final answer; these count as incorrect while retaining their retrieval and efficiency measurements.
The 36 runs contain 2{,}052 question executions, with no recorded reranker failures.

For each question, let $S$ be the distinct documents surfaced throughout the trajectory and $R$ the positively labeled documents.
Recall@all is $|S\cap R|/|R|$; Precision@all is $|S\cap R|/|S|$, or zero if $S$ is empty.
Both deduplicate documents across the trajectory and are macro-averaged over questions within each run.
Unlabeled documents count as nonrelevant.
Lightning runs switched between API providers and local serving under heavy throttling.
For Luna, each reranking configuration ran on a separate vLLM server under a different load; the resulting queuing delays changed agent-side prompt-cache reuse, so billed cost depended on serving conditions as well as trajectories.
We therefore compare search counts rather than monetary costs.
Token-priced usage is retained only for accounting in \Cref{app:cost}; local-inference estimates do not measure serving expenditure.
\Cref{tab:piika} reports six-run means and sample standard deviations.
\Cref{tab:piika-pvalues} compares \method{} with each baseline on answer accuracy, Recall@all, Precision@all, and searches, using two-sided run-level Welch tests and one pooled 16-test BY correction family across both agents (\Cref{app:stats}); bold identifies adjusted $p<0.05$.

\begin{inplacetable}
\centering
\setlength{\tabcolsep}{5pt}
\caption{\textbf{Piika: adjusted $p$-values for ACARe versus each baseline.}
BM25 API retrieval on BrowseComp-Plus$_{\mathrm{CM}}$.}
\label{tab:piika-pvalues}
\fitappendixtable{%
\begin{tabular}{@{}lllll@{}}
\toprule
Comparison & Accuracy & Recall@all & Precision@all & Searches \\
\midrule
\multicolumn{5}{@{}l}{\textit{GPT-5.6 Luna, ACARe-4B ($\mathrm{M2}$)}} \\
vs. No reranker & \textbf{0.007} & 0.847 & \textbf{0.019} & \textbf{0.014} \\
vs. Qwen3-Reranker-4B & 1.000 & 0.847 & \textbf{0.007} & 0.635 \\
\midrule
\multicolumn{5}{@{}l}{\textit{Nemotron-3.5 Lightning, ACARe-4B ($\mathrm{M3}$)}} \\
vs. No reranker & \textbf{0.016} & 0.401 & \textbf{0.024} & 1.000 \\
vs. Qwen3-Reranker-4B & 1.000 & 0.635 & 0.098 & 1.000 \\
\bottomrule
\end{tabular}}
\end{inplacetable}

\clearpage
\subsection{Base-reranker context control on BCP}\label{app:piika-context}

We test the effect of context on the unchanged Qwen3-Reranker-4B.
Nemotron-3.5 Lightning runs in Piika on 200 BCP questions, sampled with seed 42 and stratified by one-shot BM25 recall and evidence-document count.
This sample differs from Retrieval Bench's.
We will release the query IDs and sampling code.
Six runs per configuration yield 30 runs and 6{,}000 executions.

The configurations are no reranking, Qwen3-Reranker-4B in $\mathrm{M0}$, $\mathrm{M2}$, and $\mathrm{M3}$, and \method{} in $\mathrm{M3}$.
The contextual Qwen3 arms use \method{}'s prompts with unchanged weights; both $\mathrm{M3}$ arms therefore use identical context fields and prompts.
Local BM25 ($k_1=16$, $b=1$) searches the BCP corpus, returning five of 100 candidates.
Document prefixes are capped at 512 tokens for search previews, reads, and reranking; reranker preprocessing also applies the shared 300-word cap.
Lightning uses high reasoning effort with a 32K-token completion budget and a two-hour timeout.

We use the gold-answer judge and metric definitions in \Cref{app:piika-climbmix}.
Recall@all and Precision@all each average gold and evidence qrel scores equally per question.
\Cref{tab:piika-context} reports means and sample standard deviations on the shared 0--100 quality scale.
\Cref{tab:piika-context-pvalues} reports all 16 BY-adjusted, two-sided Welch tests of \method{} against four alternatives on four metrics, using six run-level means.
Baseline comparisons remain descriptive.
Monetary cost is excluded because runs switched between API providers and local serving; retained agent and judge usage is included in \Cref{app:cost}.

All 200 questions remain in every metric.
Of 231 unsuccessful executions, 174 exhaust the 32K output budget, 48 stop without a final answer, seven time out, and two encounter context-length API errors.
These receive zero accuracy while retaining retrieval and efficiency measurements.
Counts are 37, 61, 58, 51, and 24 for no reranking, Qwen3 $\mathrm{M0}$, $\mathrm{M2}$, $\mathrm{M3}$, and \method{} $\mathrm{M3}$, respectively.
Timeout counts are 1, 3, 1, 1, and 1 in that order; the API errors affect only Qwen3 $\mathrm{M2}$.

All four means improve from Qwen3 $\mathrm{M0}$ to $\mathrm{M2}$, $\mathrm{M3}$, and finally \method{} $\mathrm{M3}$.
Accuracy rises from 48.2\% to 54.8\%, 55.9\%, and 57.6\%; recall from 54.2\% to 61.2\%, 63.9\%, and 65.1\%; and precision from 14.4\% to 19.2\%, 20.1\%, and 21.2\%, while searches fall from 10.1 to 9.3, 8.7, and 8.4.
\method{} significantly improves all four metrics over Qwen3 $\mathrm{M0}$ ($p\leq0.002$), and recall, precision, and searches over $\mathrm{M2}$.
Against Qwen3 $\mathrm{M3}$, it gains 1.7 pp accuracy, 1.2 pp recall, and 1.0 pp precision, with 0.2 fewer searches; these differences are not significant after correction ($p=0.277$, $0.415$, $0.135$, and $0.890$).
Thus \method{}'s input formulation benefits the unchanged base, while the trained checkpoint achieves the best means on all four metrics.

\begin{inplacetable}
\centering
\setlength{\tabcolsep}{3pt}
\caption{\textbf{BCP context control: adjusted $p$-values} with Nemotron-3.5 Lightning and BM25 in Piika.
Bold: adjusted $p<0.05$.}
\label{tab:piika-context-pvalues}
\fitappendixtable{%
\begin{tabular}{@{}llllll@{}}
\toprule
\textbf{ACARe-4B ($\mathrm{M3}$) vs.} & \textbf{Input} & \textbf{Accuracy} & \textbf{Recall@all} & \textbf{Precision@all} & \textbf{Searches} \\
\midrule
No reranker & -- & \textbf{$<$0.001} & \textbf{$<$0.001} & \textbf{$<$0.001} & \textbf{0.002} \\
Qwen3-Reranker-4B & $q_t$ ($\mathrm{M0}$) & \textbf{$<$0.001} & \textbf{$<$0.001} & \textbf{$<$0.001} & \textbf{0.002} \\
Qwen3-Reranker-4B & $q_t,Q$ ($\mathrm{M2}$) & 0.155 & \textbf{0.002} & \textbf{0.001} & \textbf{0.049} \\
Qwen3-Reranker-4B & $q_t,\tau_t,Q$ ($\mathrm{M3}$) & 0.277 & 0.415 & 0.135 & 0.890 \\
\bottomrule
\end{tabular}}
\end{inplacetable}

\FloatBarrier

\clearpage
\subsection{Shared agent and judge prompts}
\suppressfloats[t]

\Cref{fig:piika-prompt-agent,fig:piika-prompt-judge} reproduce the shared agent-loop user prompt and gold-answer judge prompt.
The Piika judge differs from the rollout judge in \Cref{fig:rollout-prompt-judge}: it requires JSON output, handles acceptable answer alternatives and missing final answers, and does not instruct the judge to extract confidence.
Answer accuracy uses its Boolean \texttt{correct} field; missing final answers receive zero.

\begin{figure}[!htbp]
\begin{mdframed}[font=\footnotesize, roundcorner=10pt, linewidth=1pt, innerleftmargin=10pt, innerrightmargin=10pt, innertopmargin=5pt, innerbottommargin=5pt]
\setlength{\parindent}{0pt}\setlength{\parskip}{0.5\baselineskip}
You are a research and retrieval agent using only the provided tools.

Workflow:\\
1. Use search with a concise raw query string based on the original question.\\
2. Prefer short lexical searches over long natural-language rewrites.\\
3. Inspect the ranked hits returned directly by search before rewriting the query.\\
4. If a promising candidate document appears, inspect it with read\_document.\\
5. Follow the read\_document schema and any continuation guidance returned by the tool.\\
6. Use search refinements only when they add a genuinely new clue from what you already saw.\\
7. Every call to search and read\_document must include reason as the first argument. Keep it specific, under 100 words, and focused on the clue, gap, candidate, or ranking issue.\\
8. Satisfy every requested output below; use the same research pass for all outputs.\\
9. As soon as you have enough evidence and candidates, stop using tools and answer in plain assistant text.

Answer output:\\
- Produce a concise answer supported by the documents you found.\\
- Cite supporting docids inline when possible.\\
- Keep Exact Answer directly responsive to the question.

Your final response must contain exactly these sections in this order:\\
Answer:\\
Explanation: \{your explanation for your final answer. Cite supporting docids inline in square brackets [] at the end of sentences when possible, for example [123].\}\\
Exact Answer: \{your succinct, final answer\}\\
Confidence: \{your confidence score between 0\% and 100\%\}

Do not include any text before, after, or between those sections beyond the requested fields and ranked document lines.

If you later receive a user steer telling you to submit now, stop using tools immediately and answer right away with the exact final response format below. Do not do more research after that steer.

Question: \{question\}
\end{mdframed}
\caption{Piika agent-loop user prompt for the answer-generation experiments, with the benchmark question replaced by \{question\}.
All retrieval configurations share this two-tool workflow.}
\label{fig:piika-prompt-agent}
\end{figure}

\begin{figure}[!htbp]
\begin{mdframed}[font=\footnotesize, roundcorner=10pt, linewidth=1pt, innerleftmargin=10pt, innerrightmargin=10pt, innertopmargin=5pt, innerbottommargin=5pt]
\setlength{\parindent}{0pt}\setlength{\parskip}{0.5\baselineskip}
You are an evaluation judge.

Your job is to determine whether the response's final answer is semantically equivalent to the known correct answer.\\
Do not solve the question yourself.\\
Do not use outside knowledge.\\
Focus only on whether the response's final answer matches the correct answer.\\
Allow harmless wording differences, equivalent formatting, and added correct detail.\\
The correct-answer field may be a JSON array of acceptable alternatives; matching any one alternative is correct.\\
For numerical answers, allow small formatting differences and obvious equivalent forms.\\
If the response does not contain a final answer you can extract, set extracted\_final\_answer to null and correct to false.

Return exactly one JSON object and nothing else.\\
Do not wrap the JSON in markdown or code fences.\\
Use this exact schema:\\
\{\\
  "extracted\_final\_answer": string \textbar{} null,\\
  "correct\_answer": string,\\
  "reasoning": string,\\
  "correct": boolean,\\
  "confidence": number\\
\}

Requirements:\\
- confidence must be a number between 0 and 100\\
- correct must be true or false\\
- repeat the provided correct answer exactly in correct\_answer\\
- reasoning must explain only whether the extracted final answer matches the correct answer

Question: \{question\}

Response:\\
\{response\}

Correct answer: \{correct\_answer\}
\end{mdframed}
\caption{Piika reference-based answer-judge prompt, used with GPT-5.5 at low reasoning effort.
Braced input placeholders replace the question, final agent response, and reference answer.}
\label{fig:piika-prompt-judge}
\end{figure}

\FloatBarrier

\end{document}